\documentclass[12pt,preprint]{aastex}

\newcommand{\etal}{{et al.}\ }
\newcommand{\eg}{{e.g.,}\ }
\newcommand{\ie}{{i.e.,}\ }
\newcommand{\distmod}{($m-M_{\rm V}$)}
\newcommand{\Av}{$A_{\rm V}$}

\shorttitle{Precision Star Cluster Ages}
\shortauthors{von Hippel \etal}

\begin{document}

\title{Inverting Color-Magnitude Diagrams to Access Precise Star Cluster
Parameters: A Bayesian Approach}

\author{Ted von Hippel\altaffilmark{1}, William H. Jefferys\altaffilmark{1},
James Scott\altaffilmark{2}, Nathan Stein\altaffilmark{1},
D. E. Winget\altaffilmark{1}, Steven DeGennaro\altaffilmark{1},
Albert Dam\altaffilmark{3}, Elizabeth Jeffery\altaffilmark{1}}

\altaffiltext{1}{Department of Astronomy, University of Texas at Austin, 1
University Station C1400, Austin, TX 78712-0259, USA; ted@astro.as.utexas.edu}
\altaffiltext{2}{Trinity College, Cambridge CB2 1TQ, UK}
\altaffiltext{3}{Department of Computer Sciences, University of Texas at Austin,
1 University Station C0500, Austin, TX 78712-0233, USA}

\begin{abstract}
We demonstrate a new Bayesian technique to invert color-magnitude diagrams
of main sequence and white dwarf stars to reveal the underlying cluster
properties of age, distance, metallicity, and line-of-sight absorption, as
well as individual stellar masses.  The advantages our technique has over
traditional analyses of color-magnitude diagrams are objectivity,
precision, and explicit dependence on prior knowledge of cluster
parameters.  Within the confines of a given set of often-used models of
stellar evolution, the initial-final mass relation, and white dwarf
cooling, and assuming photometric errors that one could reasonably achieve
with the Hubble Space Telescope, our technique yields exceptional
precision for even modest numbers of cluster stars.  For clusters with 50
to 400 members and one to a few dozen white dwarfs, we find typical
internal errors of $\sigma$([Fe/H]) $\leq$ 0.03 dex, $\sigma$\distmod\ 
$\leq$ 0.02 mag, and $\sigma$(\Av) $\leq$ 0.01 mag.  We derive cluster
white dwarf ages with internal errors of typically only 10\% for clusters
with only three white dwarfs and almost always $\leq$ 5\% with ten white
dwarfs.  These exceptional precisions will allow us to test white dwarf
cooling models and standard stellar evolution models through observations
of white dwarfs in open and globular clusters.  
\end{abstract}


\section{Introduction}

White dwarf cooling theory currently provides the most reliable age for
the Galactic disk (Winget \etal 1987; Oswalt \etal 1996; Leggett, Ruiz, \&
Bergeron 1998; Knox, Hawkins, \& Hambly 1999), whereas main sequence
stellar evolution provides the most reliable age for the Galactic halo
(e.g., Salaris \& Weiss 2002; Krauss \& Chaboyer 2003).  In order to
understand the detailed formation sequence of the Galactic components,
as well as the local satellite galaxies, these two time scales need to be
placed on the same absolute age system.  The only current empirical
approach available to inter-calibrate these two age systems is to derive
white dwarf (WD) cooling ages and main sequence turn-off (MSTO) ages for a
number of Galactic star clusters over a wide range of ages and
metallicities.  Much of the WD age dating work has been necessarily
limited to nearby open clusters (Claver 1995; von Hippel, Gilmore, \&
Jones 1995; Richer \etal 1998; von Hippel \& Gilmore 2000; Claver \etal
2001; von Hippel 2005, hereafter paper 1) that are young or of
intermediate age, since old WDs are faint.  Hansen \etal (2002) extended
WD age studies to one globular cluster (NGC 6121 = M4).  They derived a
precise WD age, but with large systematic uncertainties due to as-yet
uncalibrated physical effects in the coolest WDs (Fontaine, Brassard, \&
Bergeron 2001).

Even though the Hubble Space Telescope may be nearing the end of its
lifetime, it has made collecting these deep observations of WDs in open
and globular clusters possible.  At least two more open clusters (NGC
2360, NGC 2660) and one more globular cluster (NGC 6397) have been
observed with HST to sufficient depth, and those results will be
forthcoming.  The large number of 8--10m telescopes now available make it
possible to observe a few more open clusters to sufficient depth for the
WD technique, and the next decade should see 20--30m telescopes, which
will make these studies substantially easier.

While the instrumentation has been improving and there has been steady
work on improving WD cooling and traditional main sequence stellar
evolutionary models, there have not been sufficient advances in the
statistical machinery available to compare star cluster observations with
those models, particularly for WDs.  In this paper we present the first
phase of our effort to develop this statistical machinery.  Specifically,
we present a new Bayesian technique that has the ability to objectively
incorporate all our prior knowledge, including stellar evolution, star
cluster properties, and data quality estimates, while comparing data for
each cluster to any available theoretical model.  We chose to employ a
Bayesian approach precisely because so much is known about stellar
evolution and star clusters, and because this approach allows us to test
how cluster properties depend on the input models or model ingredients.

The power of the Bayesian approach is impressive, and we show below both
the excellent precision one can obtain in the primary cluster parameters
(age, metallicity, distance, and reddening) and the range of related star
cluster and stellar evolution problems that can be addressed.  The goal of
this paper is to present the Bayesian technique and demonstrate its
internal precision.  In subsequent papers we will derive improved WD and
MSTO ages for clusters, with the long-term goal of intercalibrating WD and
MSTO ages up to the ages of the oldest globular clusters.

\section{Baseline Stellar Cluster Model}

We chose a single set of stellar evolution ingredients to build and test
the Bayesian approach.  We use this model set to test the sensitivity of
the derived WD and MSTO ages to the cluster parameters of [Fe/H], \Av,
distance, age, number of cluster stars, and assumed photometric error.

For our baseline Stellar Cluster Model we chose a Miller \& Scalo (1979)
initial mass function (IMF), main sequence and giant branch stellar
evolution time scales of Girardi \etal (2000), the initial (main sequence)
- final (white dwarf) mass relation of Weidemann (2000), WD cooling time
scales of Wood (1992), and WD atmosphere colors of Bergeron \etal (1995).
Using these ingredients we simulate star cluster color magnitude diagrams
(CMDs), and using Bayesian techniques discussed below, we invert cluster
CMDs to recover the probability distribution of the cluster parameters.

When simulating a cluster, each star is randomly drawn from the IMF and,
based on a user-specified binary star fraction, randomly assigned to be a
single star or a binary with a companion also randomly drawn from the
IMF.  Note that although an IMF is required to simulate a cluster, the
implied age from either the MSTO technique or the WD technique is
insensitive to the IMF.  The IMF serves only to increase or decrease the
population of stars of interest, \eg MSTO stars or WDs.  If there are
insufficient stars, particularly if the cluster is young, then the few
cluster stars coupled with the IMF can create a statistical uncertainty to
locating the MSTO or perhaps even finding WDs.  Binaries of nearly any
mass ratio have a similar effect.  WDs in binaries are generally not
recognized and MSTO stars in such systems are found brighter and generally
redder than the MSTO, and therefore they do not help define the MSTO.  For
these reasons and for simplicity in this study, we set the binary fraction
to 0\%, substantially lower than the typical value for open clusters of
$\geq30$\%.  For simplicity, we use only H-atmosphere (DA) WDs in our
present simulations.  While He-atmosphere (DB) WDs make up $\leq10$\% of
field star WD samples (7\% in Kleinman \etal 2004), to date no DBs have
been found in open clusters (Kalirai \etal 2005).  A limitation of our
cluster simulations is that stars with masses $\leq 0.25 M_\sun$ are not
included, thus producing an unrealistic lower limit to the main sequence.
Since the focus of this study is on stars that can become WDs, this
simplification is merely one of presentation.

Other stellar evolution (\eg Yi \etal 2001; Baraffe \etal 1998; Siess,
Dufour, \& Forestini 2000) and WD cooling (\eg Benvenuto \& Althaus 1999;
Hansen 1999) models could have been used, and will be added to our code
later, but for the present purposes, the above-mentioned, often-used
models adequately cover parameter space and allow us to build and test
the Bayesian machinery.

After producing simulated CMDs we incorporate realistic photometric errors
assuming reasonable cluster parameters, \eg \distmod\ = 12.5 and \Av\ = 0
to 1, and assuming observations are obtained with the HST or similar
imaging instrument able to observe to $V = 27$ with S/N = 15\footnote{From
experience, S/N=15 is required to obtain good morphological rejection of
background galaxies at HST resolution (von Hippel \& Gilmore 2000).}.  We
use a conservative upper limit to the photometric precision of S/N=200,
though we do not incorporate systematic calibration errors.  Our Stellar
Cluster Model limits are currently set by the Girardi \etal (2000) and
Wood (1992) models, and these limits are 100 Myr to 4.5 Gyr and Z =
0.0004 to 0.030 ([Fe/H] $\approx -1.676$ to +0.198).  This is adequate
parameter space for significant age and metallicity exploration and to
demonstrate the technique, though we clearly need to push the technique to
greater ages.

Our cluster simulations do not include mass segregation nor other dynamical
processes, potentially important in open clusters, especially for the
lowest mass stars; these typically have little effect on the measured WD
mass fraction (von Hippel 1998; see also Hurley \& Shara 2003 who find
that the WD luminosity function and mass function are insensitive to
dynamical effects at 0.5 to 1 half-mass radii).  Simulated clusters
specifically tuned to match real clusters using our Stellar Cluster Model
have been presented in paper 1 (figs.\ 4--10).  Here we do not attempt to
match actual clusters, \ie we do not tune distance, reddening,
metallicity, cluster richness, and age, but rather we explore hypothetical
clusters that cover the parameter space available to us.  The CMDs for two
such clusters are presented in Figures 1 and 2 for log(age) = 9.0 and 9.5,
respectively.  The masses of a few WDs from across the cooling sequence
are indicated in the second panels of both figures.  Between 1 Gyr
(Fig.\ 1) and 3.2 Gyr (Fig.\ 2) the WD terminus has evolved from $M_{\rm
V} \approx 13$ to $M_{\rm V} \approx 14.5$, and the simulated photometric
errors have increased for the faintest WDs.

In the next section, we outline the Bayesian technique that we will use in
forthcoming studies to invert actual CMDs.  In verifying the technique,
rather than apply our Bayesian code to actual clusters with necessarily
unknown parameters, we instead apply our code to simulated clusters.  The
analyses of simulated data sets test the degree to which an entirely
consistent set of stellar models, along with realistic photometric errors,
yield the original input parameters.  Our Bayesian analyses thus test the
internal precision of our technique and its sensitivity to photometric
errors, given the many non-linear aspects of stellar evolution.  Since all
stellar evolution models are imperfect, this approach provides a measure
of internal precision only, not external accuracy.  Our goal here is to
build a modeling procedure with internal uncertainty $\leq 5$\% in age,
allowing us, when we subsequently analyze real clusters with high-quality
data, to test for systematic problems in stellar models and ages of not
much more than 5\%.

\section{Bayesian Technique}

The goal of our Bayesian technique is to use information from the data and
from our prior knowledge to obtain posterior distributions on the
parameters of our model.  Our prior knowledge is encoded in prior
distributions on the model parameters.  The model parameters include
cluster parameters such as age and metallicity and an initial mass for
each cluster star.  These parameters are the inputs to our Stellar Cluster
Model, which we use to derive predicted photometric magnitudes.  The
likelihood function then compares the predicted magnitudes with the
observed (or simulated) data.  

Bayes Theorem relates the posterior distribution to the prior distribution
and the likelihood function.  If $\mbox{\boldmath{$M$}} = (M_1,M_2,\ldots,M_N)$
is a vector of initial masses of all stars in the cluster and
$\mbox{\boldmath{$\Theta$}} = (T,$ [Fe/H], \Av, \distmod) is a vector of cluster
parameters, then we can treat our Stellar Cluster Model as a function
$G(\mbox{\boldmath{$M$}},\mbox{\boldmath{$\Theta$}})$ that maps every reasonable choice
of $(\mbox{\boldmath{$M$}},\mbox{\boldmath{$\Theta$}})$ to a resultant set of
photometric magnitudes.  To obtain the likelihood, we assume that the
errors in our measurements are independently distributed and Gaussian with
known variance.  Suppose there are $N$ stars in the cluster and we have
observed them through $n$ different filters.  Then the observed data form
an $n \times N$ matrix $\mbox{\boldmath{$X$}}$ with typical element $x_{ij}$
representing the magnitude in the $i$th filter of the $j$th star. By
assumption, each magnitude is normally distributed:
\begin{equation}
x_{ij}\sim N(\mu_{ij},\sigma^2_{ij}),
\end{equation}
where $\mu_{ij}$ and $\sigma^2_{ij}$ are the mean and variance of the
modeled photometry through filter $i$ of star $j$.  The means and
variances also form $n \times N$ matrices, which we call
$\mbox{\boldmath{$\mu$}}$ and $\mbox{\boldmath{$\Sigma$}}$.  The full likelihood is then
\begin{equation}
p(\mbox{\boldmath{$X$}}|\mbox{\boldmath{$\mu$}},\mbox{\boldmath{$\Sigma$}})=\prod_{j=1}^N 
\left( \prod_{i=1}^n \left[ \frac{1}{\sqrt{2\pi\sigma_{ij}^2}}
\exp\left(\frac{-(x_{ij}-\mu_{ij})^2}{2\sigma_{ij}^2}\right) \right] \right).
\label{likelihood}
\end{equation}
The variances $\mbox{\boldmath{$\Sigma$}}$ come from our knowledge of the
precision of our observations.  The means $\mbox{\boldmath{$\mu$}}$ are the
predicted photometric magnitudes that we obtain from the Stellar Cluster
Model:
\begin{equation}
\mbox{\boldmath{$\mu$}} = G(\mbox{\boldmath{$M$}},\mbox{\boldmath{$\Theta$}}).
\end{equation}
Thus, the likelihood can be expressed in terms of the variables of our
problem and the underlying Stellar Cluster Model:
\begin{equation}
p(\mbox{\boldmath{$X$}}|\mbox{\boldmath{$\mu$}},\mbox{\boldmath{$\Sigma$}}) = 
p(\mbox{\boldmath{$X$}}|G,\mbox{\boldmath{$M$}},\mbox{\boldmath{$\Theta$}},\mbox{\boldmath{$\Sigma$}}).
\end{equation}
Computationally, (\ref{likelihood}) is the most useful form of the
likelihood because changing the underlying Stellar Cluster Model leaves
(\ref{likelihood}) unchanged.  A different Stellar Cluster Model is just a
different function, say $H$, such that
\begin{equation}
\mbox{\boldmath{$\mu$}} = H(\mbox{\boldmath{$M$}},\mbox{\boldmath{$\Theta$}})
\end{equation}
or even
\begin{equation}
\mbox{\boldmath{$\mu$}} = H(\mbox{\boldmath{$M$}},\mbox{\boldmath{$\Theta^\prime$}}).
\end{equation}
for a different set of cluster parameters $\mbox{\boldmath{$\Theta^\prime$}}$.

In Bayesian analysis, all model parameters require prior distributions.
We have tried to select priors that are consistent with astronomers'
knowledge of likely values for the various parameters.  To reflect the
fact that low mass stars are much more numerous than high mass stars and
to be consistent with our Stellar Cluster Model where we used the Miller
\& Scalo (1979) IMF, we set the prior distribution on the logarithm of a
star's mass proportional to the Gaussian distribution:
\begin{equation}
p(\log(M))\propto \exp(\frac{-(\log(M) + 1.02)^2} {0.917}),
\end{equation}
where the constants are from the fit derived by Miller \& Scalo and the
IMF is bounded at 0.15 $M_\sun$ and 100 $M_\sun$.  For metallicity,
absorption, and distance modulus we use Gaussian priors in the common
logarithm versions of these quantities ([Fe/H], \Av, \distmod).  We assume
we have reasonable knowledge of the values and uncertainties of these
parameters for a given cluster.  This knowledge should come from outside
information, not from the color-magnitude data we intend to analyze.  Our
prior on $T$, the base-10 logarithm of the cluster's age, is uniform
between $T=8.0$ and $T=9.7$, and zero elsewhere.  This is a power law
prior on the age with exponent -1, which adequately reflects the
observation that younger clusters are more common than older clusters.
Note that priors from reliable, previously-derived cluster parameters are
not required for our Bayesian approach, though they may help.  The point
is that priors encode any previously determined parameters, where they are
available.  In some cases constraining priors (\eg small $\sigma$([Fe/H])
may turn out to be required for precise results, in other cases, such as
the ones studied here, constraining priors are unnecessary for precise
results.

Given the prior distributions and the likelihood, we obtain the posterior
distributions of the parameters from Bayes theorem, which states that the
posterior density $p(\theta|y)$ on model parameters $\theta$ given data
$y$ is
\begin{equation}
p(\theta|y) = \frac{p(y|\theta) p(\theta)}{p(y)},
\end{equation}
where $p(y|\theta)$ is the likelihood and $p(\theta)$ is the prior density
on the model parameter $\theta$.  The denominator, $p(y)$, is obtained by
integrating the numerator over all possible values of $\theta$ so that
\begin{equation}
p(\theta|y) = \frac{p(y|\theta) p(\theta)}{\int p(y|\theta) p(\theta) d\theta}.
\end{equation}

In our problem, it is impossible to compute the integral $\int p(y|\theta)
p(\theta) d\theta$ analytically. Instead, we use Markov chain Monte Carlo
(MCMC) to approximate the posterior distribution (Casella \& George 1992;
Chib \& Greenberg 1995).  The MCMC algorithm allows us to generate a
sample from the posterior distribution.  We construct a Markov chain such
that once it has converged, results of each iteration of the algorithm are
distributed approximately according to the posterior distribution, and we
regard the history of the chain as a random sample from the posterior. We
can thus obtain quantities of interest, such as sample means, without
having to analytically compute the normalized posterior distribution.

Our analysis relies on the Metropolis-Hastings algorithm (Chib \&
Greenberg 1995), which proceeds as follows:  Suppose the current state at
iteration $t$ is $\theta^t=\theta$.  Propose to move to some new state
$\theta^\star$.  This proposal is generated with density
$q(\theta^\star|\theta)$.  Compute the Metropolis-Hastings factor
\begin{equation}
\alpha =
\min\left[\frac{p(\theta^\star|y)q(\theta|\theta^\star)}{p(\theta|y)q(\theta^\star|\theta)},1\right]
\end{equation}
and set $\theta^{t+1}=\theta^\star$ with probability $\alpha$.  Otherwise,
set $\theta^{t+1}=\theta$.  Our sample is the parameter sequence
$(\theta^n,\theta^{n+1},\ldots,\theta^N)$ where $N$ is the total number of
iterations and $n$ is the number of iterations before the chain
converges.  We discard the first $n-1$ iterations, which are referred to
as the burn-in.  Note the advantage of this method: since $\int
p(y|\theta) p(\theta) d\theta = \int p(y|\theta^\star) p(\theta^\star)
d\theta^\star$,
\begin{equation}
\alpha = \min\left[\frac{p(y|\theta^\star)p(\theta^\star)q(\theta|\theta^\star)}
{p(y|\theta)p(\theta)q(\theta^\star|\theta)},1\right].
\label{MHfactor}
\end{equation}
We can compute everything in (\ref{MHfactor}) without calculating any
intractable integrals.

The efficiency of the Metropolis-Hastings algorithm depends heavily on the
choice of proposal distribution $q$. A common choice is a symmetric
distribution centered at the current value. This is the ``random walk"
Metropolis-Hastings sampler. This method has the advantages of simplicity
and ease of implementation. However, the sampler can be inefficient if the
distribution's width is inappropriate---the sampler might propose
excessively small steps and take too long to traverse parameter space, or
it might propose unreasonably large jumps and frequently reject steps.
Another option is to choose a proposal distribution that approximates the
posterior distribution. This kind of sampler is known as an
``independence" sampler since $q(\theta^\star|\theta)=q(\theta^\star)$, so
that each proposed value is independent of the current state. The closer
the proposal distribution approximates the target distribution, the higher
the acceptance rate and (generally speaking) the more efficient the
sampler.

\subsection{MCMC Sampling}

One of the chief problems in designing the MCMC sampler was overcoming the
strong correlations between many of the variables.  For instance, for a
given position on the CMD, an increase in the age of the cluster will
require a decrease in the mass of a WD, and vice versa.  Since each
parameter is sampled on individually in sequence, without removing the
correlations the sampler can only take small steps in age or mass; if too
large a step is taken, the proposed star's photometry will be too far from
the observed position and the step will be rejected.  In a similar way,
metallicity and distance modulus are correlated with each main sequence
star's mass, with each other, and with reddening.  While from a
theoretical standpoint removing these correlations is not required to
obtain valid results, the number of iterations needed to be certain the
entire posterior distribution is well sampled would necessitate far more
computation time than is practical.

Fortunately, over the ranges that our MCMC typically samples, these
correlations are all nearly linear.  To remove the WD age-mass
correlation, we introduce a new parameter, $U$, and a constant, $\beta$,
defined by:
\begin{equation}
M_k = \beta (T_k - \overline{T}) + U_k,
\end{equation}
where $M_k$, $U_k$, and $T_k$ are the mass, decorrelated mass parameter,
and logarithm of the cluster age at the kth iteration, respectively, and
$\overline{T}$ is the mean log cluster age.  Then, rather than sampling on
mass, we sample on $U$ for each star.  The MCMC algorithm then computes
the mass at each iteration from the above equation.  Figures 3 and 4
present the log(age) sampling history before and after correlation is
removed, within the same MCMC run.  In Figure 3, age values spanning
$\sim100$ iterations are correlated, meaning little new is learned about
the posterior distribution within that correlation length.  In Figure 4,
the log(age) history is well-sampled and each iteration usefully samples
the posterior distribution.  The new parameter, $U$, is then decorrelated
from distance modulus and metallicity in a similar manner.  Finally, the
distance modulus and metallicity are decorrelated from one another and
then from reddening.

In order to improve the efficiency of our MCMC algorithm, we still need to
address several sampling issues.  For some parameters, the correlations
become non-linear, often at their extreme values.  For other parameters,
the correlations consists of two or more separate, nearly linear, pieces
with different slopes.  For the brightest (youngest) WDs, the correlations
between mass and age can be incredibly tight, and further work needs to be
done for these objects to more precisely trace these correlations.

The burn-in period for our MCMC runs consisted of a brief (5000 samples)
period to settle close to the correct values and adjust step sizes.  This
was followed by two periods of 5000 samples each to calculate the
correlation between mass and age for WDs, two more to calculate the
correlations between modulus and mass for main sequence stars, and between
modulus and absorption, and two more to calculate the metallicity-main
sequence mass and metallicity-absorption correlations.  Finally, there is
another 5000 sample period to adjust step sizes again.  The whole burn-in,
except for the initial settling-in period, is then repeated to more
precisely determine the correlation factors, for a total burn-in period of
70,000 samples.

\section{Demonstration and Discussion}

For the tests presented here we placed priors on cluster distance moduli,
metallicities, and absorption values.  The priors were normal
distributions centered on the simulated Stellar Cluster Model parameters
for \distmod, [Fe/H], and \Av, with the further requirement that \Av\
$\geq$ 0.  The \Av\ = 0 runs are limiting cases, and for these we assumed
$\sigma$(\Av) = 0, \ie there was no sampling on \Av.  For the \Av\ = 0.1
and 0.3 cases, we assumed $\sigma$(\Av) = 0.1 and for the \Av\ = 1 case we
assumed $\sigma$(\Av) = 0.3.  For the other cluster parameters we assumed
$\sigma$([Fe/H]) = 0.3 dex and $\sigma$\distmod\ = 0.2.  All of these
priors represent conservative uncertainties for well-observed,
low-reddening clusters.  We also placed a prior on the mass distribution
for any given star with a form, as discussed above, based on the low-mass
IMF.

We simulated $B$, $V$, and $I$ photometry for clusters for the range of
parameters log(age) = 8.3, 8.7, 9.0, 9.3, and 9.5; [Fe/H] = $-$1.0,
$-$0.15, 0.0, and +0.15; N (number of cluster stars fainter than the MSTO,
including WDs) = 50, 100, 200, and 400; \distmod\ = 12.5; and \Av\ = 0,
0.1, 0.3, and 1.  Since our goal is to test the age sensitivity of the
WDs, we removed all MSTO, subgiant, and giant branch stars from each
simulation, so that what remains are WDs and essentially unevolved main
sequence stars.

While we astronomers are most comfortable studying star clusters in the
CMD, our Bayesian technique does not use the CMD, with its correlated
errors between the x-axis and y-axis, but rather it uses an
$n$-dimensional space, where $n$ is the number of filters available and
the units are magnitudes.  In the numerical experiments we present here,
$n$=3, as we use $B$, $V$, and $I$ photometry.  The input to the MCMC
routine sees the CMD of Fig.\ 1, for example, in a form more akin to
Figure 5, although offset by the simulated distance modulus.  For
presentation purposes, we reduced three dimensions to two by plotting
either the $B$ or $I$ absolute magnitudes on the horizontal axis.  One
disadvantage of this plot is the large dynamic range in both axes.  Still,
the main CMD features can be discerned, \eg WDs are clearly visible in the
faintest corner of the plot.  Reddening vectors for \Av\ = 1.0 are also
shown, as is the effect of increasing distance modulus by 1.0 mag.  Both
distance and the reddening vectors are nearly parallel to the main
sequence, especially the $BV$ main sequence.  Decreasing metallicity from
[Fe/H] = 0.0 to $-0.1$ moves the main sequences in almost the opposite
directions as the reddening vector.  While there are some features in this
diagram, and while the various distance, reddening, and metallicity
vectors are not absolutely parallel and therefore not entirely degenerate,
this diagram suppresses subtleties that primarily affect stellar color.

Although we can simulate clusters younger than log(age) = 8.3, the MCMC
technique requires sampling an age range, and for younger clusters this
would often hit our (current) lower age limit of log(age) = 8.0.  For the
\Av\ cases, three simulated clusters were run for each unique set of
cluster parameters.  Any two clusters with identical parameters will yield
different CMDs as both the IMF and the simulated photometric error
distribution are sampled anew.  After creating a cluster, we pass it to
the MCMC routine with estimates of the mass of each star and estimates of
the cluster parameters as starting points.  (Our experiments show that as
long as the MCMC algorithm converges, the results do not depend on the
starting points.  Starting points within a factor of $\sim2$ in age or
metallicity, for instance, are adequate for convergence.)  Because the
MCMC sampling is still often correlated, we sample for $10^6$ iterations,
reading out every tenth value for each stellar mass and for the cluster
parameters.  Many of our MCMC runs have a correlation length of $\leq10$,
and this produces uncorrelated parameter values.  For those cases that
still remain correlated, and guided by the rule-of-thumb that one
typically wants $10^4$ uncorrelated iterations in order to adequately
sample the posterior distribution, we find $10^6$ iterations works well
for most of our simulated clusters.

Figure 6 presents a well-sampled, typical history plot of cluster age for
the cluster of Fig.\ 1.  Figures 7 and 8 present the companion history
plots for cluster [Fe/H] and \distmod.  There is a small amount of
sticking in the sampling of these two variables at the beginning of the
sequence, just after burn-in, and again near iteration 1.47 $\times 10^5$,
but otherwise these history plots are well sampled.  Since there is no \Av\ 
sampling in the \Av\ = 0 case of Fig.\ 1, Figure 9 presents the \Av\ history
plot for a cluster with the same parameters, except input \Av\ = 1.
Histograms of these four types of history plots (Figure 10) are the
estimates of the posterior probability distributions.  In Figure 10 we
present also the \Av\ = 0.3 and 1 cases.  Figure 10 shows that the
posterior distributions of log(age) and the other cluster parameters are
close to normal, and furthermore that changing the absorption causes no
strong bias in the results (more on absorption and bias below).  From
these posterior probabilities we can calculate statistics of interest, \eg
mean, median, $\sigma$, percentiles, etc., and these are presented,
below.  Figure 11 presents the posterior probabilities of mass for four
stars from the cluster simulation of Fig.\ 1.  The first panel shows the
mass posterior for a high mass WD, the second for a lower mass WD, the
third for a main sequence star not far below the turn-off, and the final
panel presents a low mass main sequence star.  In all cases the mass
distribution is nearly centered on the input mass--the mass value before
its photometry was subject to random error--except in the case of the
lowest mass star, where it differs by only 0.002 $M_\sun$.  For the main
sequence stars, the mass distribution is particularly narrow, showing that
within the assumption of a specific model, precise photometry yields
precise masses.  The WD mass distributions are slightly broader because we
plot the zero-age main sequence (ZAMS) masses for these stars and a wider
range of initial main sequence masses is converted into a narrow range of
WD masses via the initial-final mass relation (Weidemann 2000).

In Figure 12, we check the differences between the mean log(age) values of
the distributions and the input log(age) values compared to the standard
deviations of the posterior log(age) distributions ($\sigma$).  We are
essentially asking what the deviation of each result is in units of its
standard deviation.  For the \Av\ = 0 case, the distribution of errors is
very similar to the overplotted normal distribution.  Formally, the error
distribution is also close to normal with average = 0.037, median = 0.055,
standard deviation = 0.985, and skew = $-0.191$.  This comparison is a
sanity check on the self-consistency of our implementation of the Bayesian
technique and whether the standard deviation statistic adequately captures
the shapes of the posterior distributions.  For the \Av\ = 0.1, 0.3, and 1
cases, our MCMC approach tends to be biased high in age by 0.56~$\sigma$,
0.26~$\sigma$, and 0.12~$\sigma$, and for the higher absorption cases to
have a pronounced non-normal distribution.  These distributions are
virtually identical if we plot median values instead of means of the
posterior distributions.  While we are still trying to understand some of
the subtleties of the higher \Av\ cases, these offsets, which corresponds
to 2.6\%, 1.2\%, and 0.6\% systematic errors in age, are small enough that
we set aside their resolution for now.  Figure 12 demonstrates,
particularly for low absorption values, that the standard deviations in
the posterior distributions are accurate assessments of the uncertainties
due to photometric errors or any effects due to low number statistics,
such as having very few WDs in young or sparse clusters.

In Figure 13, we present the standard deviation log(age) uncertainty for
each model with one or more WD versus the input log(age) of the cluster.
The standard deviations are always small, typically $\leq 0.04$ dex,
corresponding to relative errors typically $\leq 10$\%, for all ages
tested.  These errors do not depend significantly on the cluster age.  In
fact, the apparent slight dependence on age seen in Figure 13 is a
combination of two other effects: there are fewer WDs in the youngest
clusters and the coolest WDs in the oldest clusters are fainter, and
therefore have higher photometric errors.  Figure 14, which plots the same
standard deviation uncertainties, now versus log($N_{\rm WD}$), shows the
most important factor in this technique--the number of WDs per cluster.
Although every WD contains age information, the quality of that
information is not the same for all WDs.  Photometry for the coolest WDs
in any cluster provides the most information (see below), yet photometric
precision drops with decreasing WD luminosity.  Therefore the age
precision does not improve as the square root of the number of WDs, but
somewhat more slowly.  Nonetheless, even with 10 WDs the statistical
(internal) error is almost always $\leq$ 0.02 dex, or $\leq$ 5\%.  Even
with three WDs, 10\% precision is usually achieved.

Since the precision is so high via this technique, it is worth taking a
small detour into the details of the age information locked up in each
WD.  Figure 15 presents the relationship between possible masses and
possible ages for six of the nine WDs of Fig.\ 1 (log(age) = 9.0).  Two
WDs are not presented because they are so close in mass to other WDs that
they crowd the figure without presenting any new information.  The
faintest WD of Fig.\ 1, has been dropped also, since its ZAMS mass (7.6
$M_\sun$) is beyond the 7.0 $M_\sun$ limit of the Weidemann (2000)
initial-final mass relation and its WD mass (1.131 $M_\sun$) is beyond the
Wood (1992) 1.0 $M_\sun$ cooling WD model limit.  Although this
extrapolated WD has properties that are internally consistent enough for
the MCMC runs, they are not pedagogically helpful in understanding the
precision in the WD technique.  Figure 15 shows that the age-mass
relationship for the hotter, brighter WDs is highly correlated, which is
the cause of the numerical difficulties with correlated sampling mentioned
above.  Is there useful age information in these hot, rapidly cooling WDs;
information that is unexploited in the traditional approach, which uses
the coolest WDs to derive a cluster's age?  

Figure 16 presents the age sensitivity of this technique\footnote{Note
that the slight changes of slope in the mass-age relationship for the
lowest mass WD is a numerical artifact cause by our use of linear
interpolation among the Girardi \etal (2000) models.  Besides being a
small effect and outside of the actual fit presented in Fig.\ 15,
choppyness due to linear interpolation serves to occasionally slightly
decrease the precision of our technique.  Higher-order interpolations have
not yet been necessary and would nearly double the run-time of our MCMC
code.} by presenting the allowed age-mass relationship for each of the WDs
of Fig.\ 15, {\it if each were the only WD in the otherwise identical
cluster.}  Running the same cluster with only one WD yields the age
constraints from an individual WD, while still relying on the cluster main
sequence to constrain the combination of metallicity, distance, and
reddening.  Now it is clear that the higher mass WDs provide the tightest
age constraints, eliminating significant age range allowed by the lowest
mass, hottest WD, for example.  Yet, in this case even the second hottest
WD contain significant age information.  Our technique can be pushed to
the point where ages can be derived for clusters without observing the
coolest WDs, and a companion paper will explore the sensitivity of that
approach (Jeffery \etal 2006).  The higher mass WDs have a much flatter
slope in the mass-age diagram since large changes in ZAMS mass do not
appreciably change the contribution of precursor time scales, nor do they
evolve at a much different rate as WDs, at least not in this age range.

Figure 16 still begs the question: Why is there any age sensitivity when
there is only one cluster WD?  The short answer to this is that the WD
region of the CMD is not highly degenerate.  Though it may be possible to
make a highly degenerate CMD with a combination of few cluster stars, high
and uncertain reddening, uncertain metallicity, etc., generally this is
not the case.  There are also other constraints on the WD properties.  A
WD cannot have a mass higher than the upper limit for creating WDs (8
$M_\sun$ in all our simulations, most likely 7--9 $M_\sun$) and a WD
cannot have a mass so low that stars with lower initial masses are still
present on the main sequence.  Changes in mass move a WD in the CMD along
essentially the same vector as changes in age for hot WDs, where precursor
ages are important, and in a perpendicular direction to age for cooler
WDs, where precursor ages are unimportant.  Figure 17 attempts to make
this clear by plotting a small portion of the CMD of Fig.\ 1 around the
simulated WDs.  The simulated WDs are presented as error bars, whereas the
input values, before photometric scatter was added, are presented as
filled circles.  Here all cluster WDs, except the highest mass (7.6
$M_\sun$) simulated WD, are plotted.  The small `+' symbols connected by
lines show the effect of holding mass constant while changing log(age) by
$\pm$ 0.01 dex, or in the case of the two highest mass WDs, by changing
log(age) by $\pm$ 0.02 dex.  Open squares show the effect of keeping
log(age) = 9.0 while adjusting the ZAMS masses by $\pm$ 2\%, or for the
two highest mass WDs, by $\pm$ 5\%.  WD isochrones for log(age) = 8.9,
8.95, 9.0, 9.05, and 9.1 are overplotted.  Ultimately, the Bayesian
technique is so sensitive because minor changes in WD mass or cluster age
of just a few percent move the expected location of any WD significantly
in the CMD.  Also, some types of photometric error, \eg the
$\sim$~2~$\sigma$ $B-V$ color error of one of the WDs in the middle of the
cooling sequence, cannot be matched by any realistic adjustment of cluster
or stellar parameters, and thus this photometric error does not drive the
fit for this WD.  In Figures 15 and 16, for example, this WD is the object
plotted third from the right.  Its ZAMS mass (3.285 $M_\sun$) and age
(log(age)=9.0) sit right in the center of the sampled values, so this
color error had no meaningful effect.  Errors in color could cause one to
mistake a WD for a field star, however.  The solution to this problem in
real clusters with possibly contaminating field stars is better photometry
or classification-level spectroscopy to confirm the WD.

Besides deriving precise ages, the Bayesian technique also can derive
precise values for the cluster parameters of metallicity, \distmod, and
\Av.  In all these cases the standard deviations in the posterior
distributions are small, typically $\leq$ 0.03 dex, $\leq$ 0.02 mag, and
$\leq$ 0.01 mag, respectively.  All of these posterior uncertainties are
an order of magnitude smaller than width of the prior distributions (0.3
dex, 0.2 mag, and 0.1--0.3 mag, respectively), demonstrating that high
quality priors in these parameters are not generally needed, at least for
low-to-moderate, single-valued absorption, and that the results are
insensitive to the exact assumed starting values for these parameters.
The cluster photometry contains a wealth of information, and the Bayesian
technique, along with an assumed model set, brings this out to high
precision.  For clusters with ten WDs, the age precision is typically
better than 5\%, easily meeting our needs for a precise statistical tool.

\section{Bayes Meets Star Clusters: Other Uses}

In our work to date, we have focused on cluster ages, particularly via the
WD technique.  Age via the MSTO technique will be next.  Our MCMC code
also derives values for the other major cluster parameters: metallicity,
distance modulus, and line-of-sight absorption, along with the individual
stellar property of mass.  For our own purposes, we intend to upgrade our
MCMC code to include binaries drawn from realistic mass ratio
distributions, field stars, DB (He atmosphere) WDs, and a wider range of
standard stellar and WD evolution models, including ages up to globular
cluster values.

Here are a few further example uses for our Bayesian technique:

1. In our effort to improve WD and MSTO ages we will systematically study
main sequence and WD model parameters that affect ages, such as core
convection prescriptions, non-standard elemental abundances, and diffusion
in main sequence stars, as well as surface convection prescriptions and
C/O phase separation in cool WDs.  

2. We intend to study the sensitivity in the implied underlying parameters
of simulated and actual clusters to the initial-final mass relation as
well as the upper mass limit for creating WDs.

3. Our code derives mass posterior distributions for every object in the
cluster.  These mass estimates would be a good starting point for IMF
studies, particularly since one can adjust the priors on mass to reflect
different assumed IMFs and one can incorporate as input any stellar
evolution model.  By adjusting the prior on the IMF, one could see how
many cluster stars are required before the resulting IMF is no longer
sensitive to the prior.

4. Once we add binaries to the MCMC code, we intend to study cluster
binaries, their masses, and mass ratios.  This would be a step forward
from the typical approach of estimating cluster binary contributions by
visually studying the distribution of stars above the single-star main
sequence.  Additional information, such as the probability of cluster
membership from proper motion or radial velocity studies, could also be
incorporated in the binary studies.

We expect to make our code publicly available within a year, after it
passes out of its development stage.

\section{Conclusions}

We have demonstrated a new Bayesian technique to invert color-magnitude
diagrams to reveal the underlying cluster properties of age, distance,
metallicity, and line-of-sight absorption, as well as individual stellar
masses.  We do not fit cluster fiducial sequences nor do we create plots
with many combinations of cluster parameters and then try to derive the
best parameters via chi-by-eye.  The Bayesian technique delivers not just
parameters and error estimates, but entire posterior distributions.
Posterior distributions for the parameters of interest are particularly
valuable when they may be non-normal, as may occur with all the coupled,
non-linear aspects of stellar evolution.  Despite the potential for
complex error distributions, we find posterior age distributions that are
close to normal in log(age).  Some other distributions, \eg some mass
distributions, are clearly non-normal.

Within the confines of a given set of often-used models of stellar
evolution, the initial-final mass relation, and WD cooling, and assuming
photometric errors that one could reasonably achieve with HST, we find
that our technique yields exceptional precision for even modest numbers of
cluster stars.  For clusters with 50 to 400 members and one to a few dozen
WDs, we find typical internal errors of $\sigma$([Fe/H]) $\leq$ 0.03 dex,
$\sigma$\distmod\ $\leq$ 0.02 mag, and $\sigma$(\Av) $\leq$ 0.01 mag.
The parameter we are most concerned with, cluster WD age, has an internal
error of typically only 0.04 dex (10\%) for clusters with only three WDs
and almost always $\leq$ 0.02 dex ($\leq$ 5\%) with ten WDs.  All of these
results have posterior distributions an order of magnitude narrower than
the priors we applied, and therefore represent the actual information in
cluster CMDs.  Cluster photometry clearly contains a wealth of
information, much of it coupled in a non-linear fashion, and the Bayesian
technique, along with an assumed model set, brings this out to high
precision.

\acknowledgments
We thank the referee, Gordon Drukier, for helpful suggestions that
improved our manuscript.  This material is based upon work supported by
the National Aeronautics and Space Administration under Grant
No.\ NAG5-13070 issued through the Office of Space Science.  We also
gratefully acknowledge a grant to JS from the NSF-funded Vertical
InteGration of Research and Education (VIGRE) program awarded by UT's
Department of Mathematics.


\figcaption{({\it a}) $BV$ and ({\it b}) $VI$ CMDs in the dereddened,
absolute magnitude plane for a representative, log(age) = 9.0 cluster with
[Fe/H] = 0.0, N = 100 main sequence plus WD stars, and photometric errors
appropriate for \distmod\ = 12.5, \Av\ = 0.  Photometric errors in the WD
region are similar in the $BV$ and $VI$ CMDs for these simulations, though
there is a x-axis scale change.  Representative ZAMS masses for the WDs
are given in solar units.}

\figcaption{Same as Fig.\ 1, except for log(age) = 9.5.  The oldest WDs
are now fainter and have larger simulated photometric errors.}

\figcaption{Correlated age vs.\ ZAMS mass sampling for a WD during
burn-in.}

\figcaption{The WD in Fig.\ 3, after age-mass decorrelation.}

\figcaption{The Bayesian code's version of the Fig.\ 1 CMDs.  Filled
squares represent the $BV$ plane and open triangles represent the $VI$
plane for solar metallicity stars.  Cluster WDs are in the upper right,
with the $VI$ sequence slightly above the $BV$ sequence.  The effects of
metallicity are shown by the open circles, which are the $BV$ and $VI$
main sequences, for [Fe/H] = $-1$.  The offsets of distance and reddening
are removed from this plot for presentation purposes.  Reddening vectors
for \Av\ = 1 for both planes are in the upper left, as is a vector
showing the effect of increasing the cluster's distance modulus by 1 mag.
The reddening vectors and distance offset vector are replotted near the
main sequences to facilitate comparison.}

\figcaption{Typical history plot of cluster log(age), here for the 1 Gyr
cluster of Fig.\ 1.  In this case, the sampling was excellent.  For
clarity, only every 100th point is plotted, and only after the initial,
70,000 iteration burn-in period.}

\figcaption{Similar to Fig.\ 6, but for cluster [Fe/H].  This particular MCMC
run had very mild correlation in [Fe/H].}

\figcaption{Similar to Fig.\ 6, but for \distmod.}

\figcaption{Similar to Fig.\ 6, but for \Av\ in a cluster with \Av\ = 1.}

\figcaption{Histograms of the MCMC history plots, for ({\it a}) log(age)
of Fig.\ 6, ({\it b}) [Fe/H] of Fig.\ 7, ({\it c}) \distmod\ of Fig.\ 8,
and ({\it d}), \Av\ of Fig.\ 9.  For the first three panels the posterior
distributions for \Av\ = 0, 0.3, and 1 are plotted and for the final panel
the latter two distributions are plotted, since there was no sampling on
\Av\ for the \Av\ = 0 case.  Every tenth iteration, which is the frequency
with which these MCMC runs were read out, is incorporated into the
histograms.}

\figcaption{Posterior probabilities of ZAMS masses for four stars from the
cluster presented in Fig.\ 1.  Panel ({\it a}) shows the mass posterior
for a high mass WD, panel ({\it b}) for a lower mass WD, panel ({\it c})
for a main sequence star not far below the MSTO, and panel ({\it d}) for a
low mass main sequence star.  In all cases the means of the mass
distributions are similar to the input masses, labeled with small vertical
marks at the bottom of each panel.}

\figcaption{The differences between the mean ages of the posterior
distributions and the input ages normalized by the standard deviations
($\sigma$) of each posterior age distribution for the ({\it a}) \Av\ = 0,
({\it b}) \Av\ = 0.1, ({\it c}) \Av\ = 0.3, and ({\it d}) \Av\ = 1 cases.
The distribution of errors is very similar to the overplotted normal
distribution for the \Av\ = 0 case and becomes subtly biased to +0.12 to
+0.56~$\sigma$ for higher values of \Av.}

\figcaption{Derived standard deviations for each model with one or more WD
versus the actual (input) ages of the clusters.  The standard deviations
are always small, typically $\leq 0.04$, corresponding to relative errors
typically $\leq 10$\%, for all ages tested.  One result (among 319 runs)
with $\sigma$(log age) = 0.105 at log(age) = 8.7 and $N_{\rm WD}$ = 1, is
not plotted for presentation purposes.}

\figcaption{Standard deviation uncertainties versus log($N_{\rm WD}$) for
the same data as presented in Fig.\ 13.  The precision of the age fit
improves approximately as the log of the number of WDs, which is an
important factor under the observer's control.}

\figcaption{WD ZAMS masses versus cluster age for six of the nine WDs for
the cluster in Fig.\ 1.  The other three WDs are not plotted for clarity.}

\figcaption{WD ZAMS masses versus cluster age for six modified versions of
the cluster presented in Fig.\ 1.  In order to explore the mass-age
correlations and see which WDs provide the greatest age constraints, nine
clusters, each with only one WD from the original nine cluster WDs, were
created.  Again, only six of these mass-age relations are plotted for
clarity.  The lowest mass WDs have the tightest mass-age correlations,
which creates greater MCMC sampling challenges.  The higher mass WDs
provide tighter age constraints.  The kinks in the lowest mass
relationship occur at boundaries of the main sequence (Girardi \etal 2000)
tracks, and are numerical artifacts.}

\figcaption{WD regions of the Fig.\ 1 CMDs.  The input WDs are plotted as
filled circles and the scattered photometry are plotted as 1~$\sigma$ error
bars.  The highest mass WD is not plotted (see text).  The `+' symbols
connected by lines show the effect of changing log(age) by $\pm$ 0.01 dex,
or in the case of the two highest mass WDs, by $\pm$ 0.02 dex.  Open
squares show the effect changing ZAMS masses by $\pm$ 2\%, or for the two
highest mass WDs, by $\pm$ 5\%.  WD isochrones for log(age) = 8.9, 8.95,
9.0, 9.05, and 9.1 are overplotted.  The reddening vectors for \Av\ = 0.1
are also shown.}

\clearpage

\begin{figure}[!t]
\epsscale{.80}
\plotone{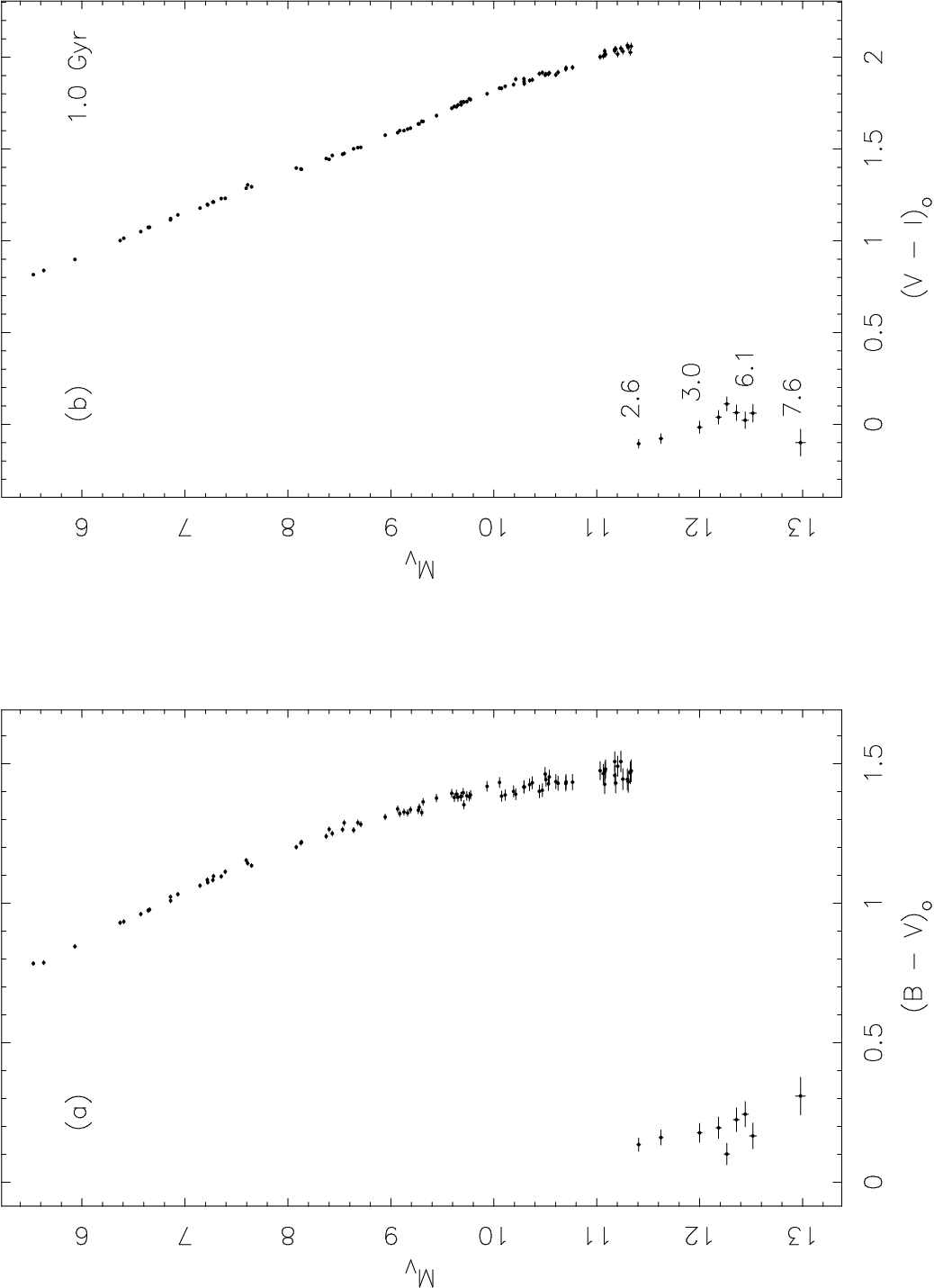}
\centerline{f1.eps}
\end{figure}
\clearpage

\begin{figure}[!t]
\plotone{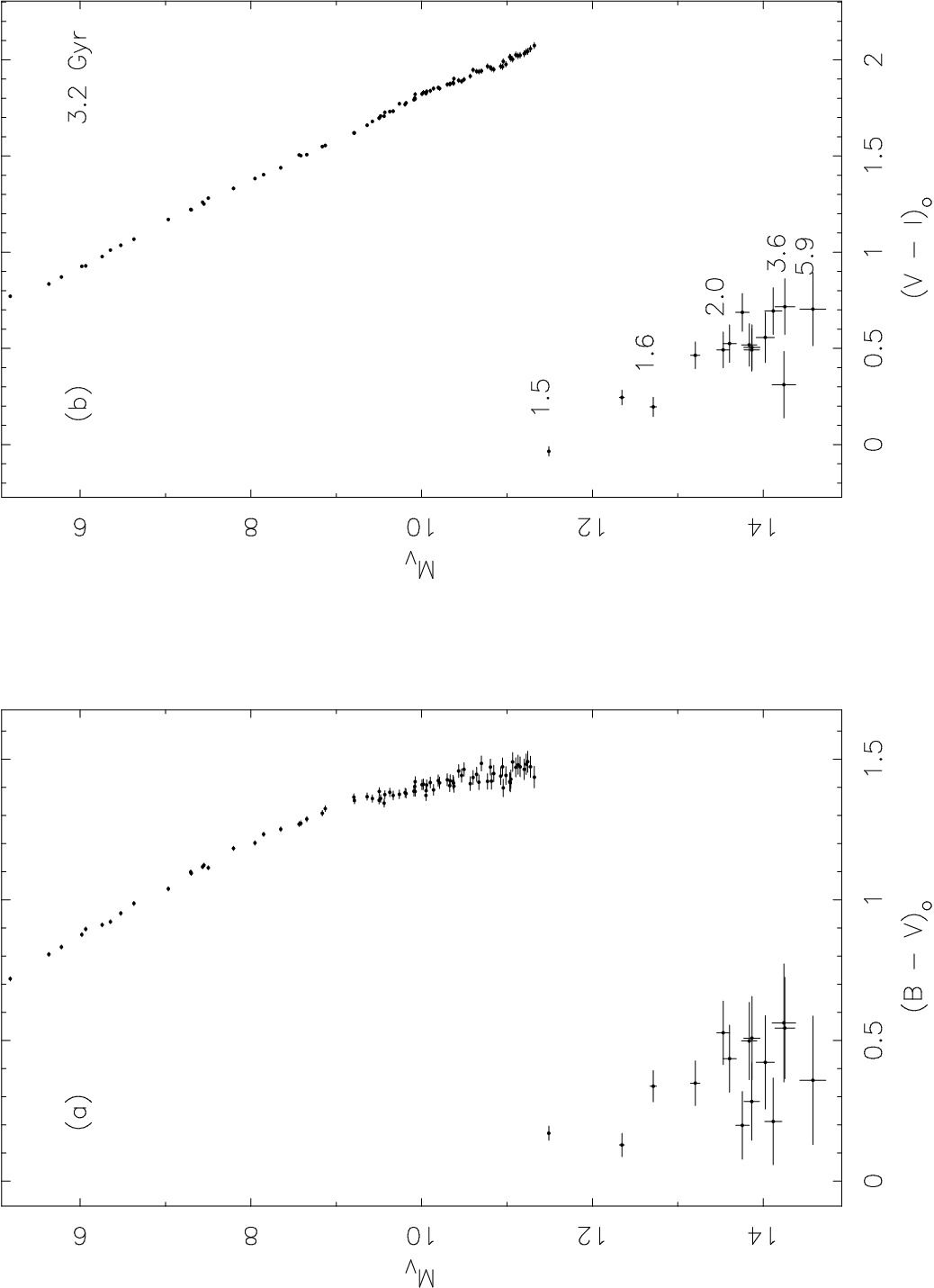}
\centerline{f2.eps}
\end{figure}
\clearpage

\begin{figure}[!t]
\plotone{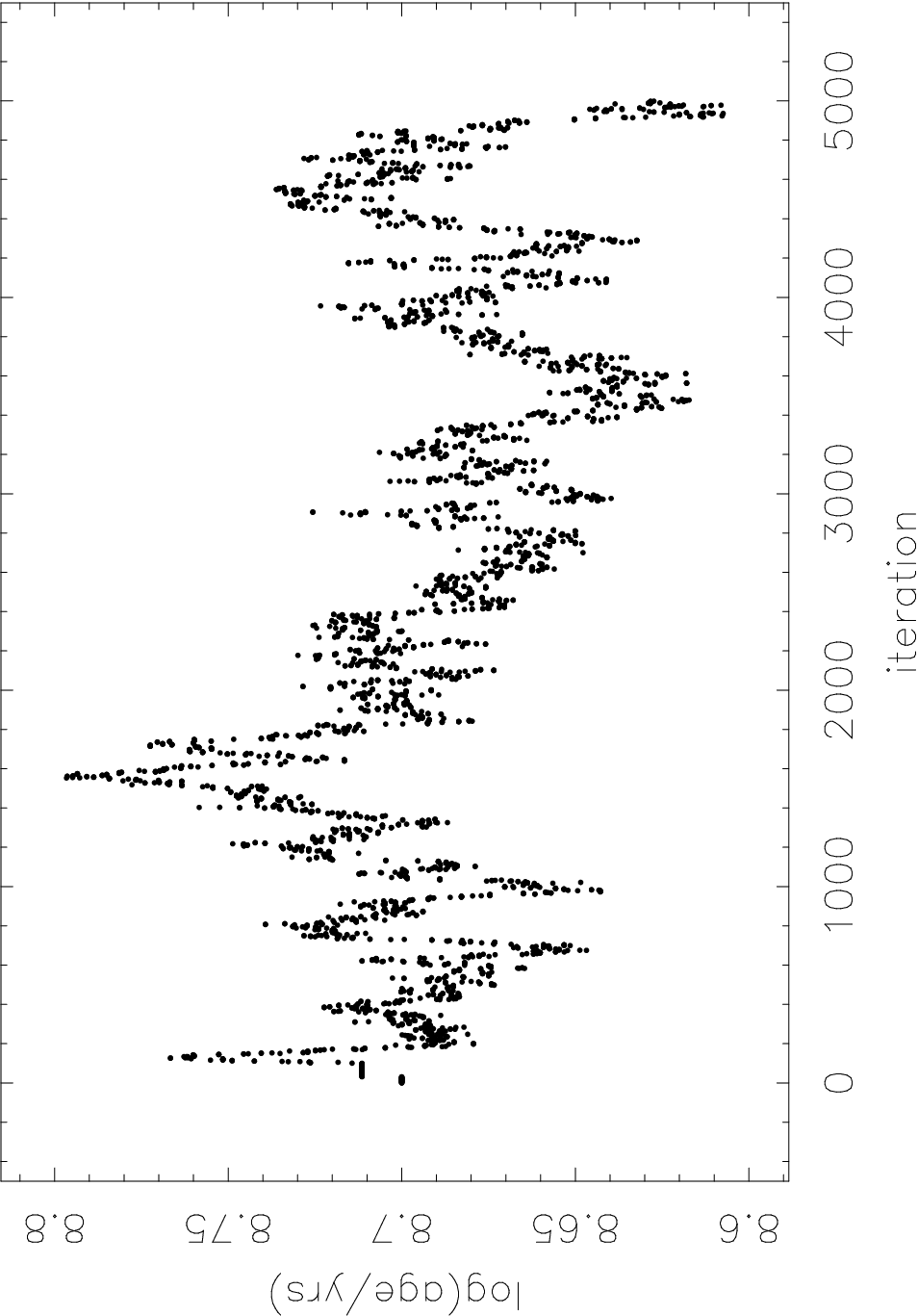}
\centerline{f3.eps}
\end{figure}
\clearpage

\begin{figure}[!t]
\plotone{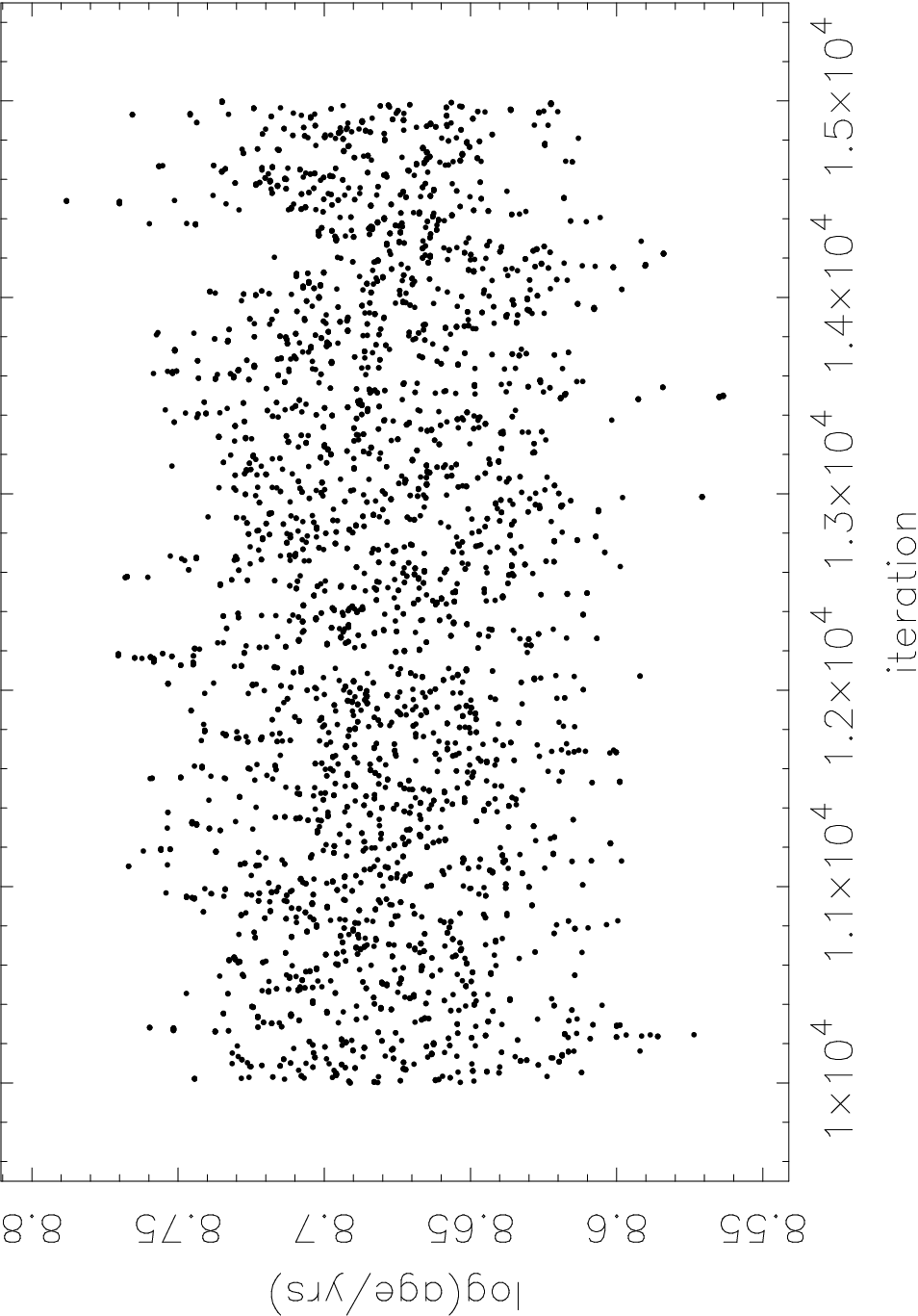}
\centerline{f4.eps}
\end{figure}
\clearpage

\begin{figure}[!t]
\plotone{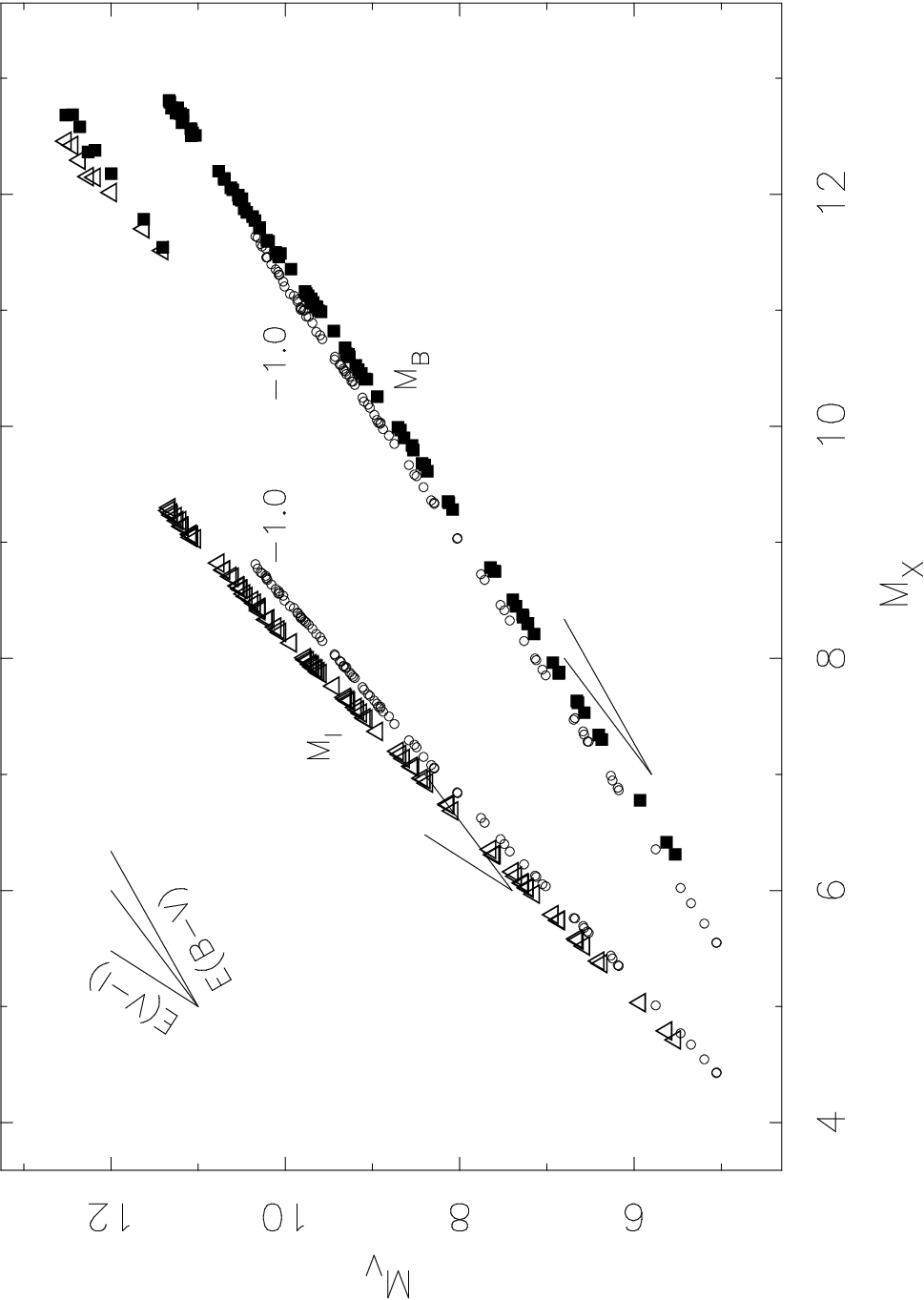}
\centerline{f5.eps}
\end{figure}
\clearpage

\begin{figure}[!t]
\plotone{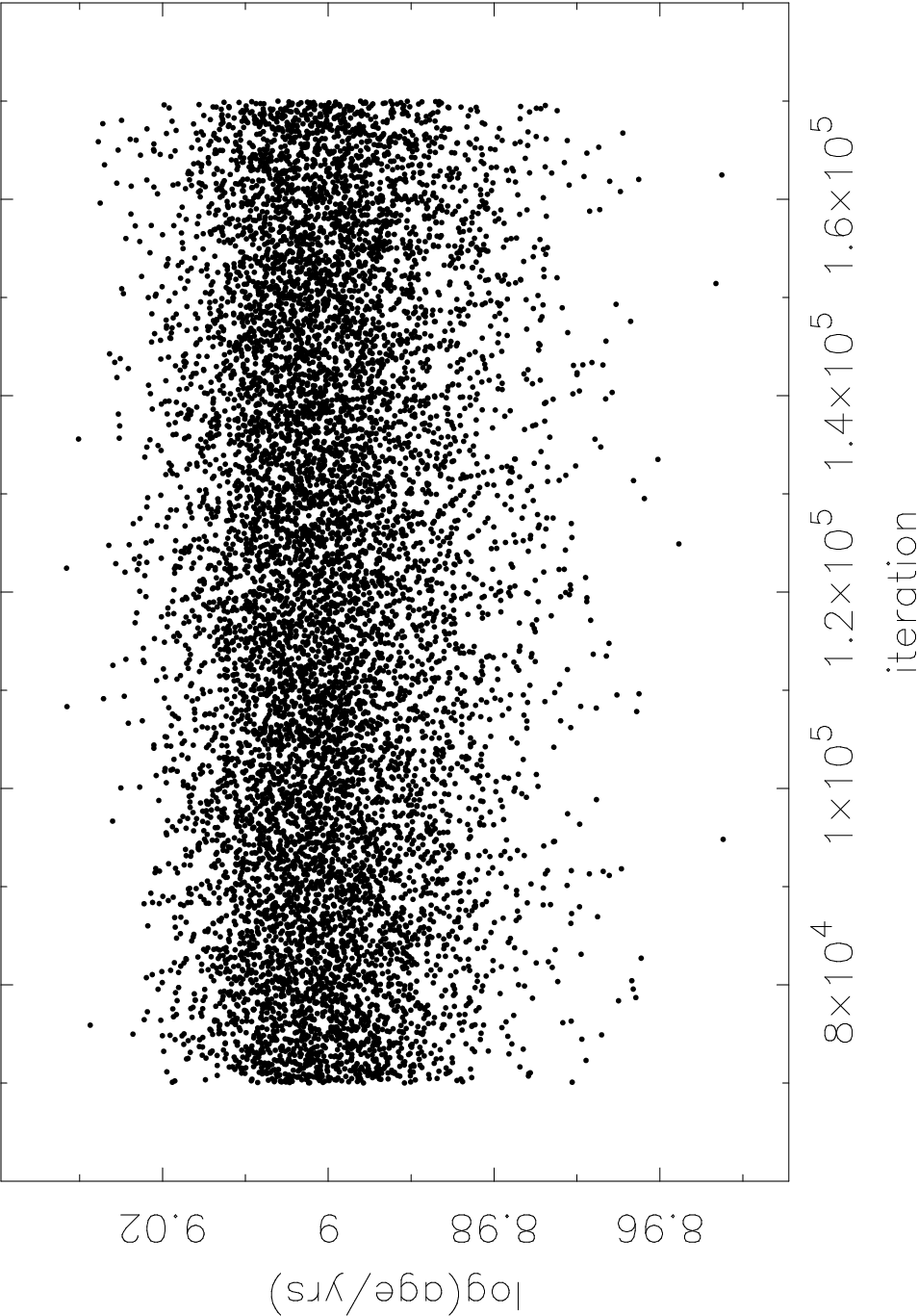}
\centerline{f6.eps}
\end{figure}
\clearpage

\begin{figure}[!t]
\plotone{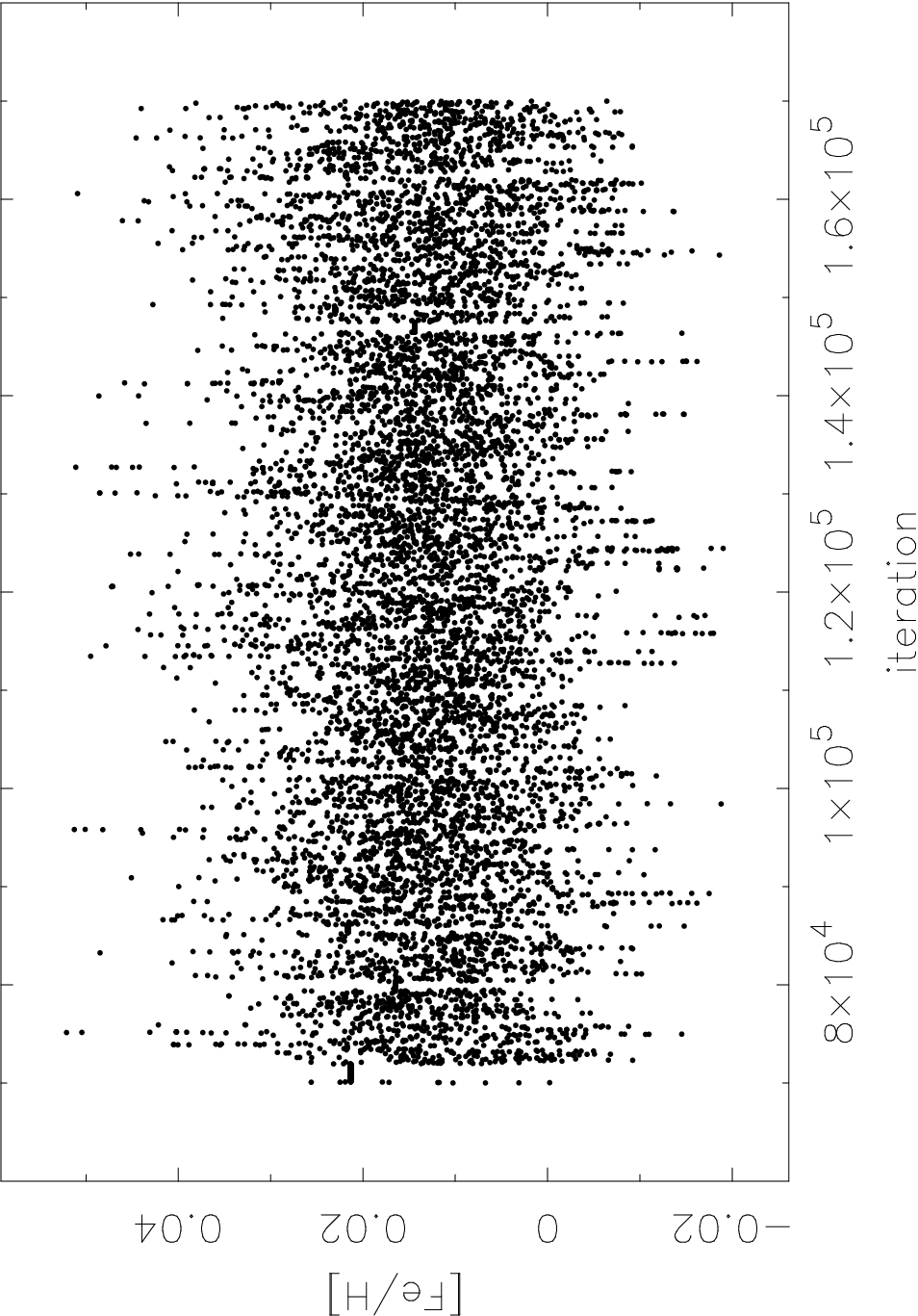}
\centerline{f7.eps}
\end{figure}
\clearpage

\begin{figure}[!t]
\plotone{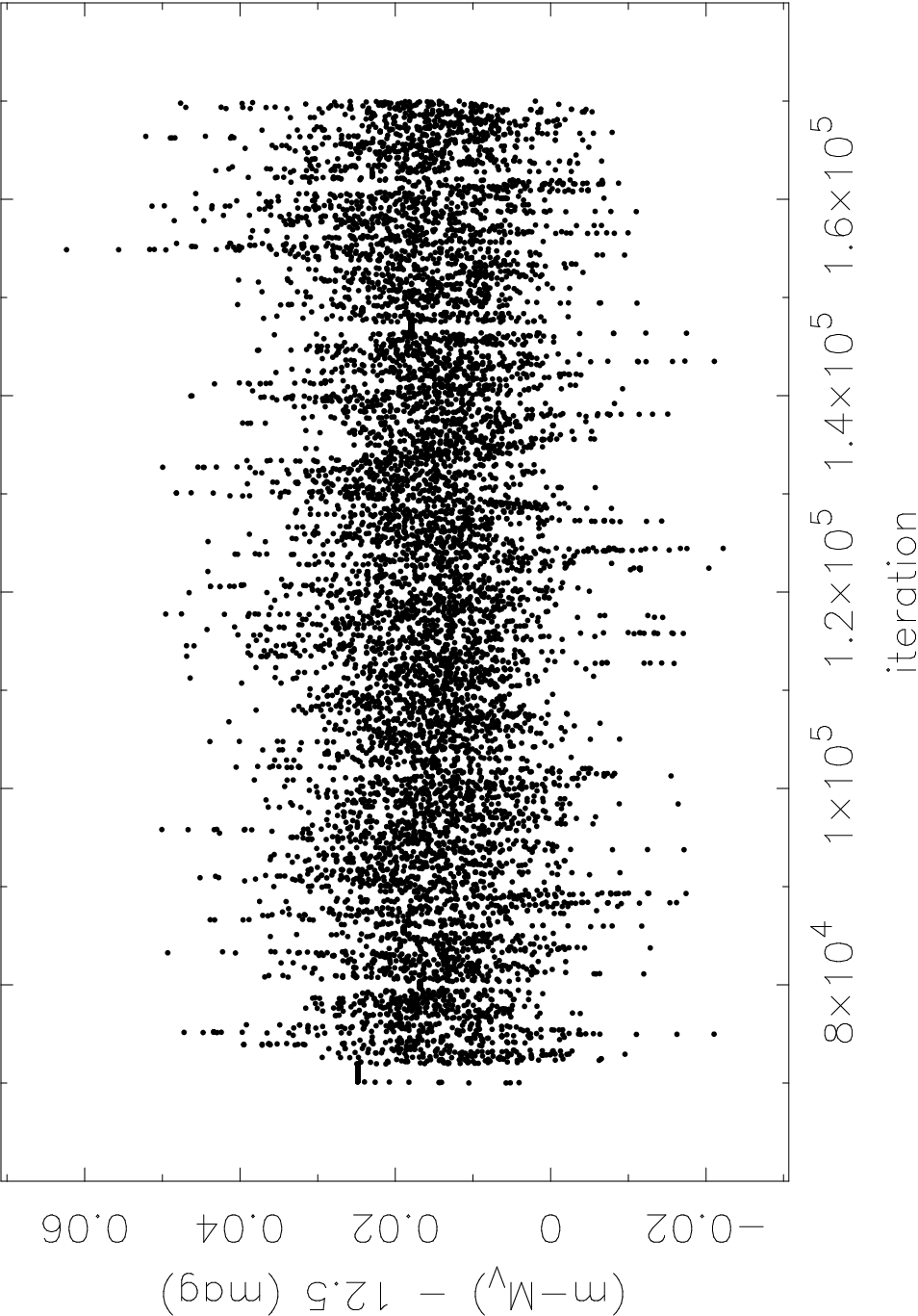}
\centerline{f8.eps}
\end{figure}
\clearpage

\begin{figure}[!t]
\plotone{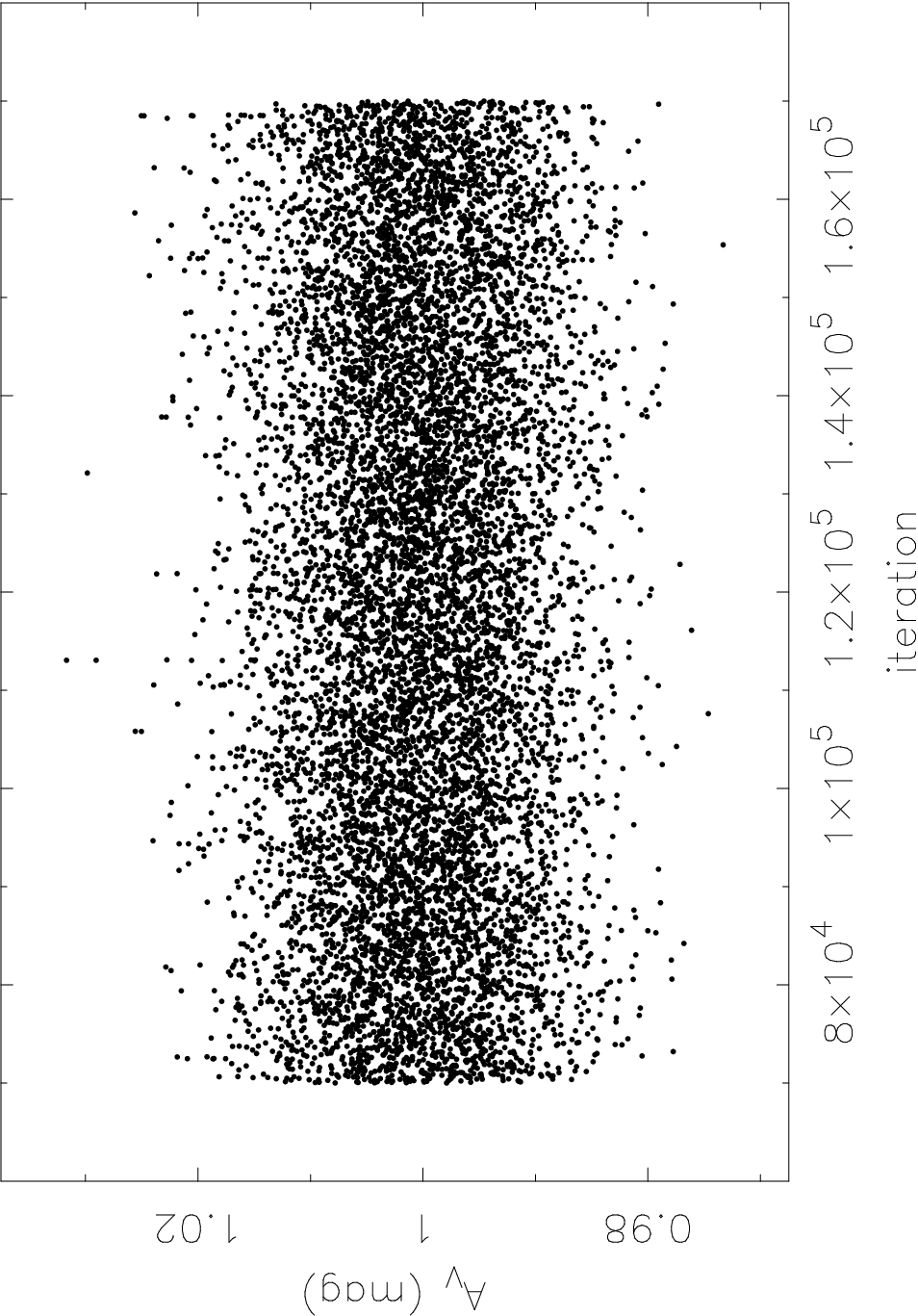}
\centerline{f9.eps}
\end{figure}
\clearpage

\begin{figure}[!t]
\plotone{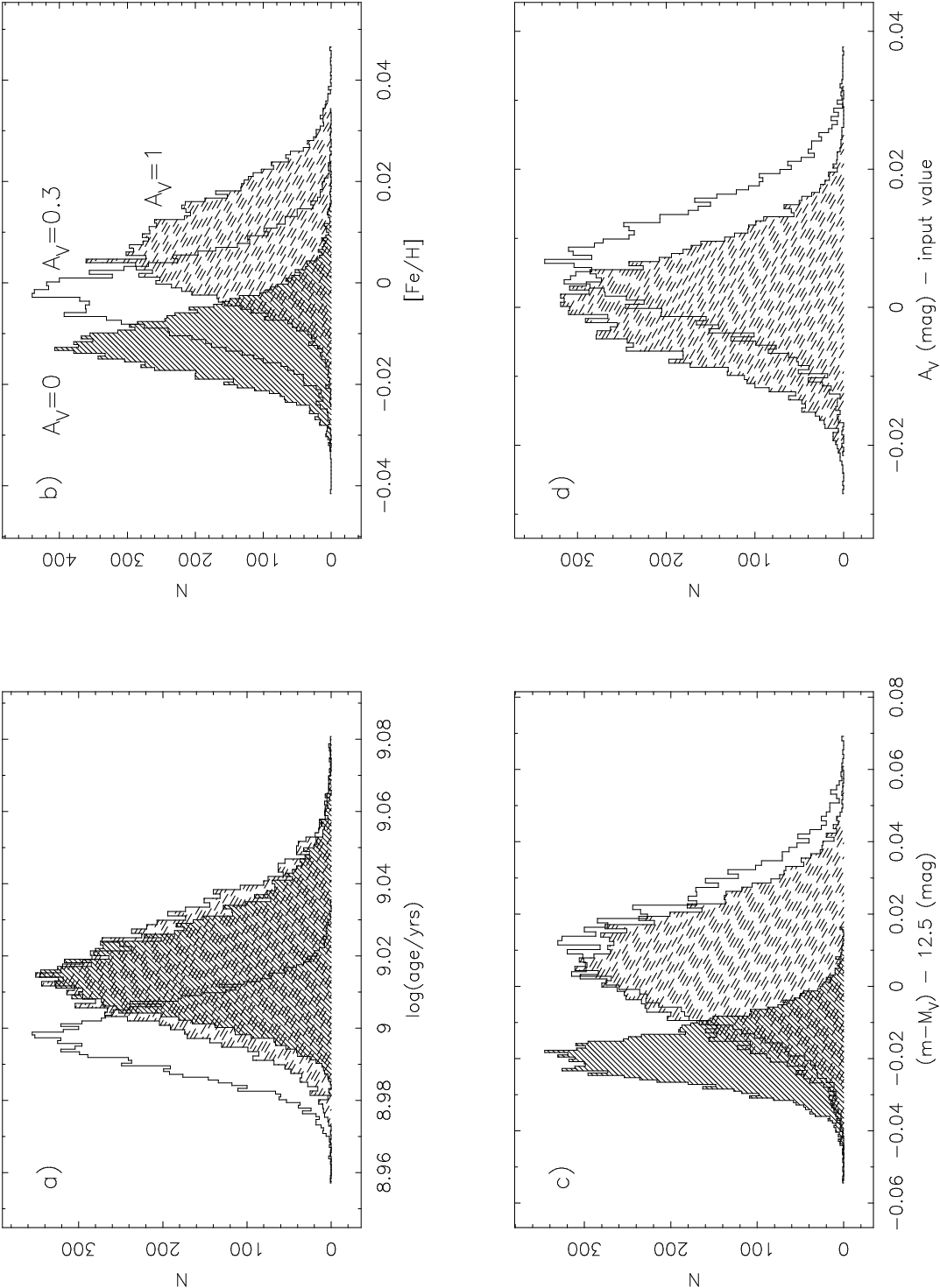}
\centerline{f10.eps}
\end{figure}
\clearpage

%
%
%

\begin{figure}[!t]
\plotone{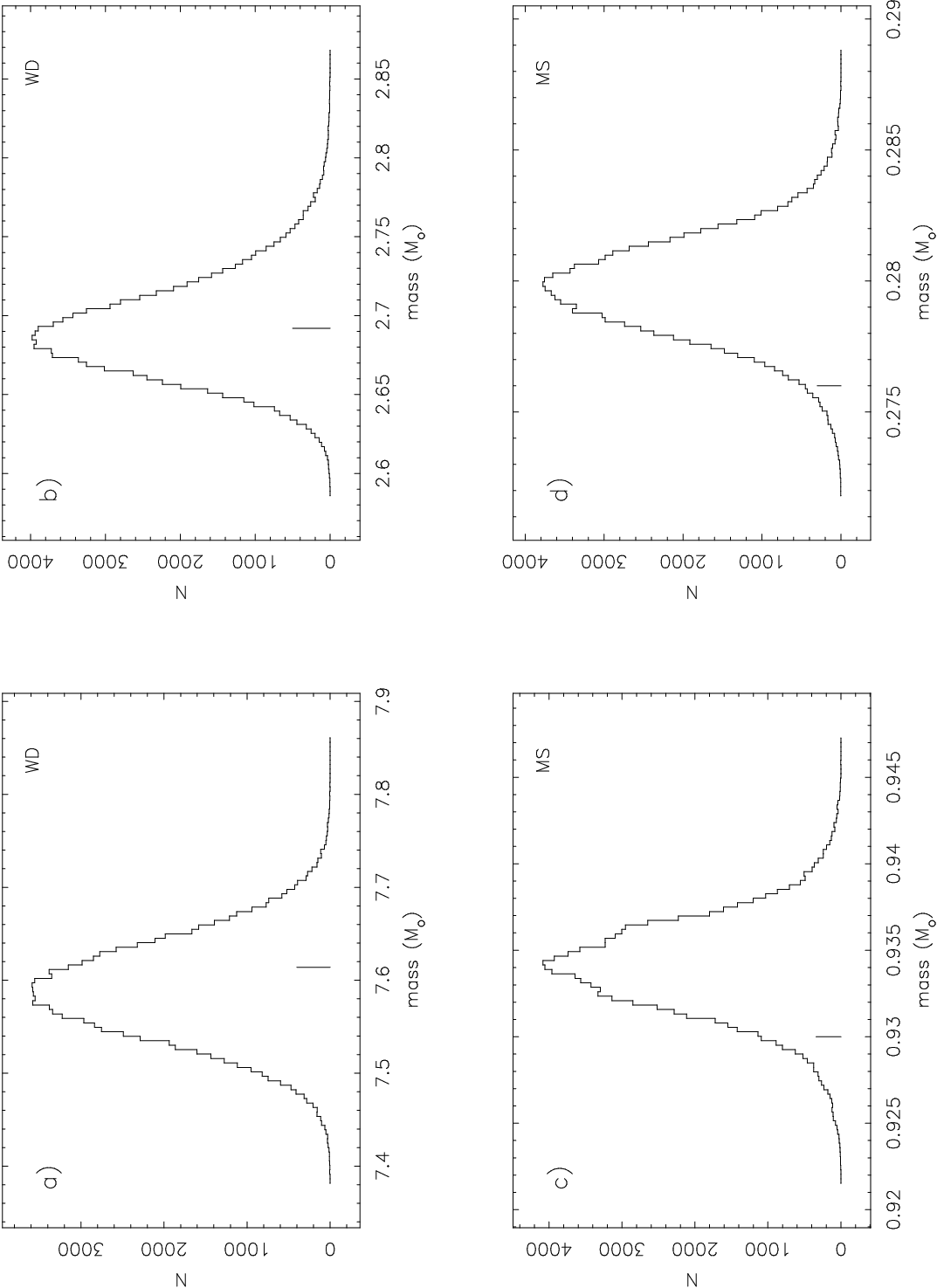}
\centerline{f11.eps}
\end{figure}
\clearpage

\begin{figure}[!t]
\plotone{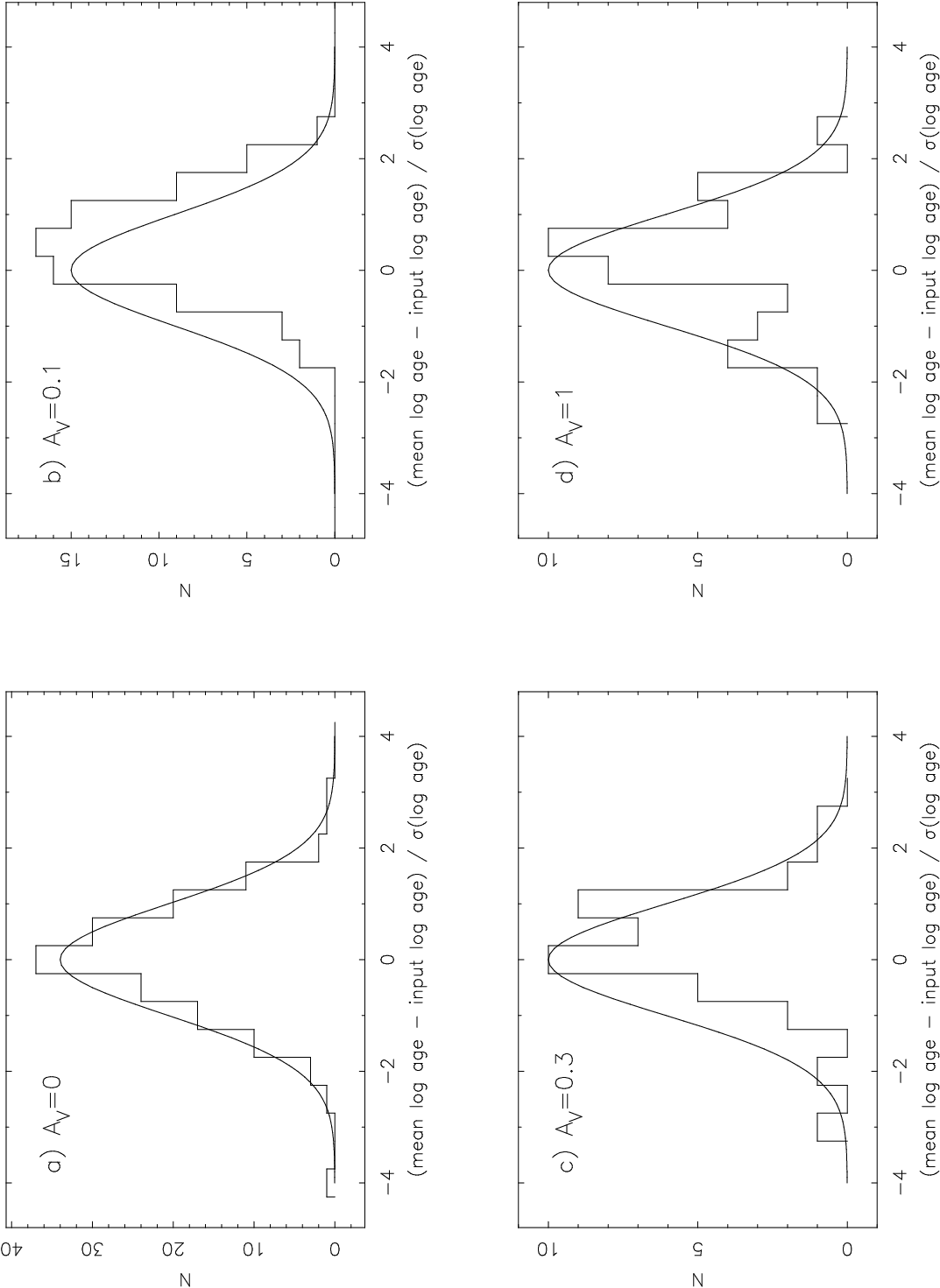}
\centerline{f12.eps}
\end{figure}
\clearpage

\begin{figure}[!t]
\plotone{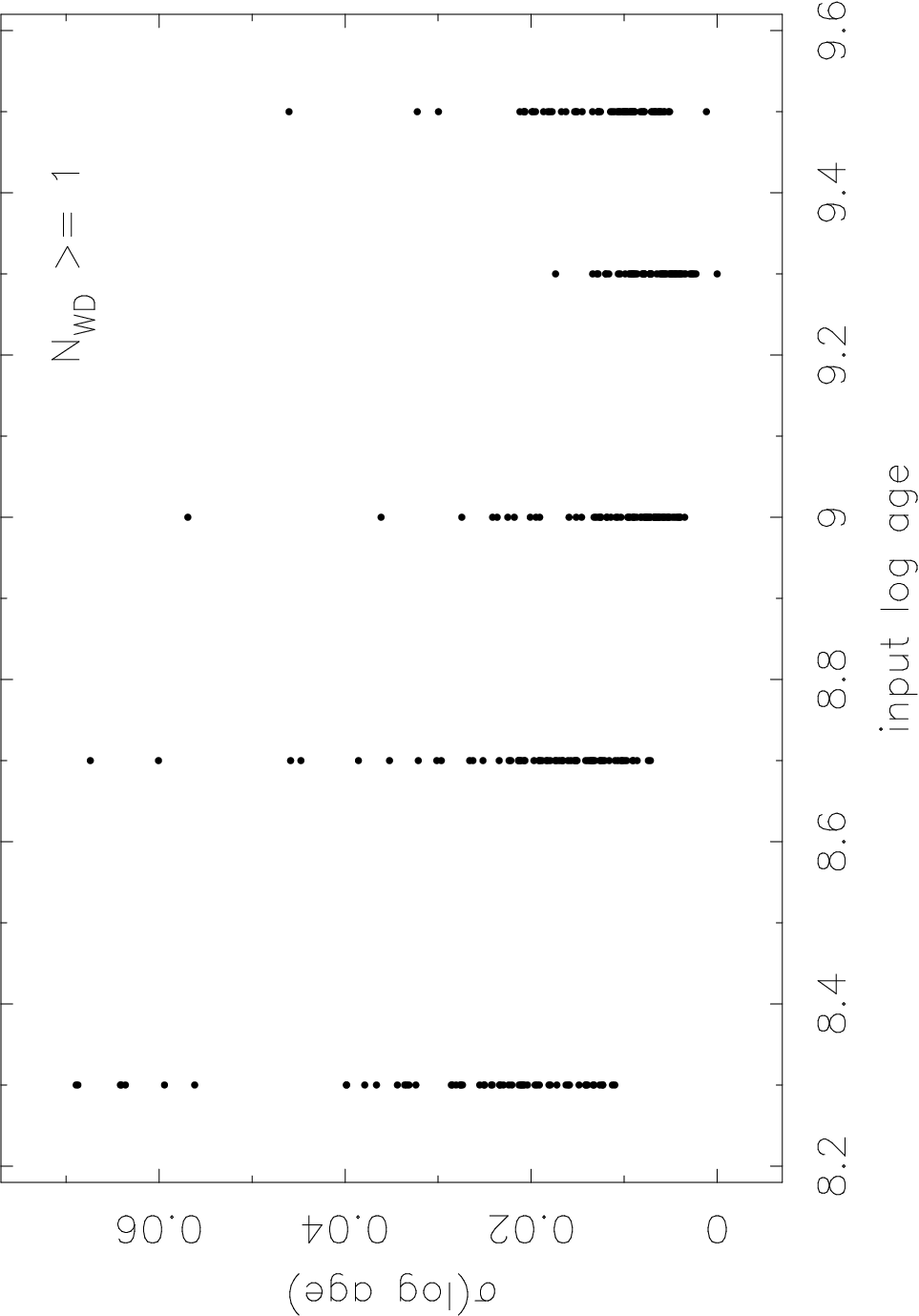}
\centerline{f13.eps}
\end{figure}
\clearpage

\begin{figure}[!t]
\plotone{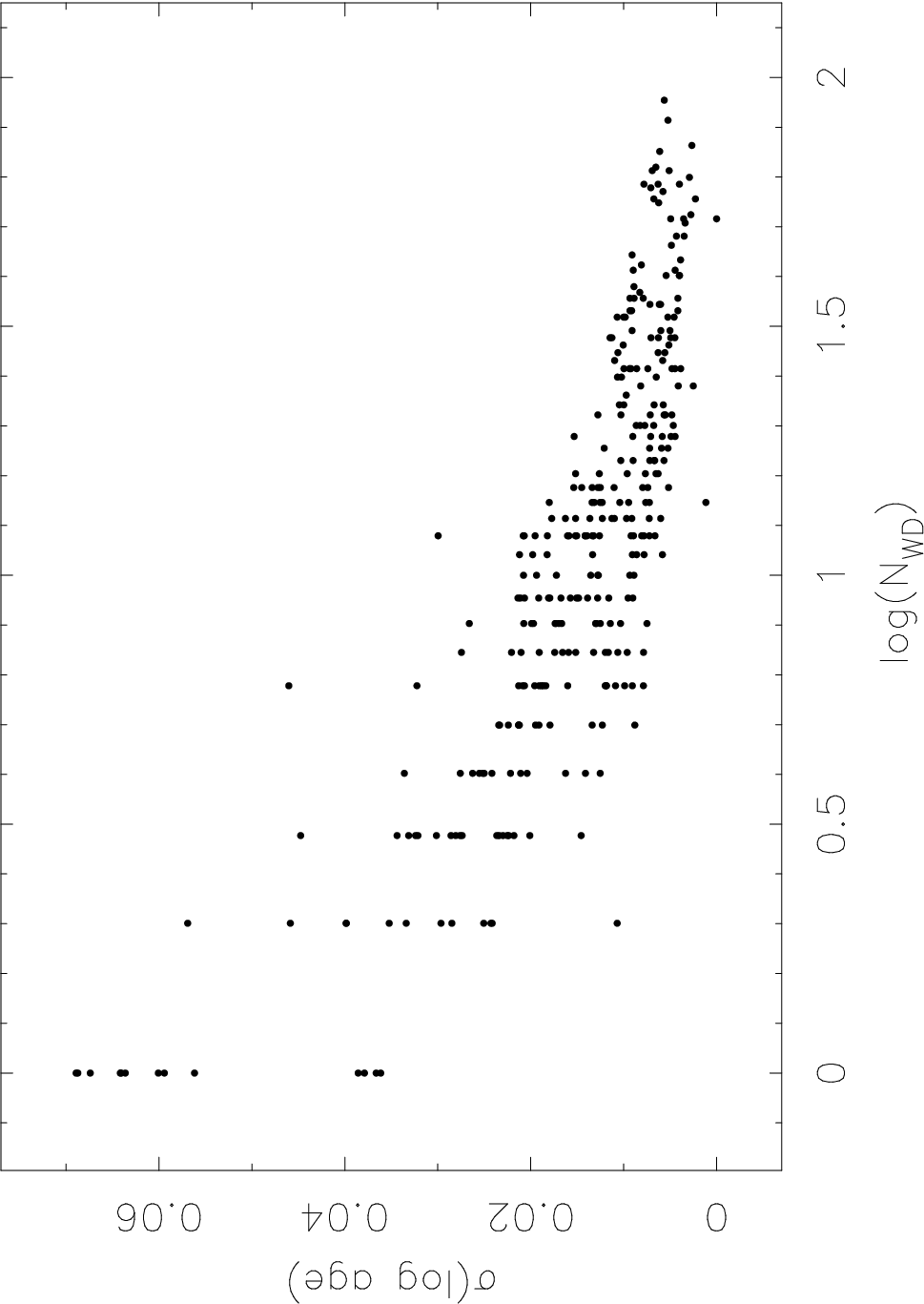}
\centerline{f14.eps}
\end{figure}
\clearpage

\begin{figure}[!t]
\plotone{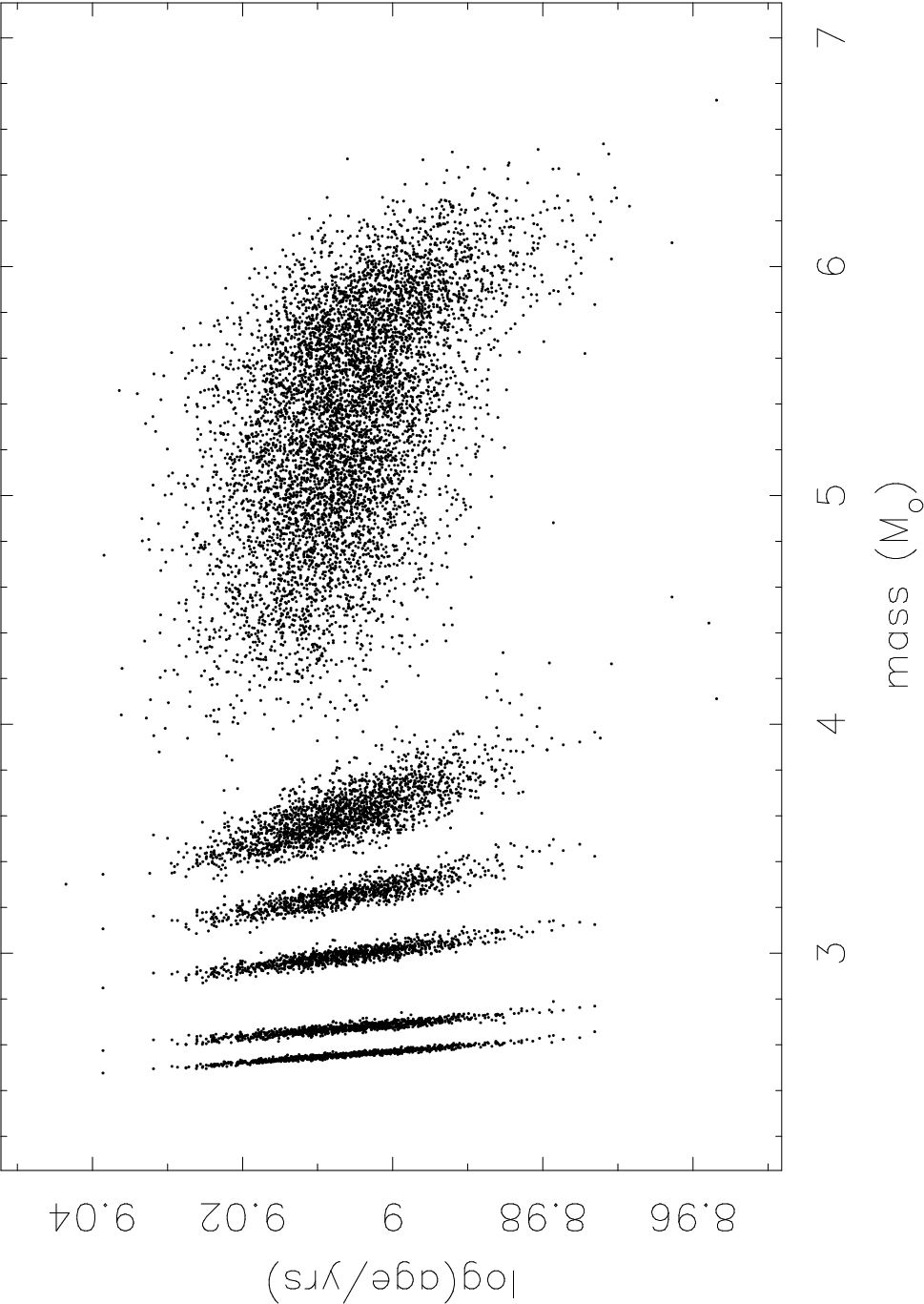}
\centerline{f15.eps}
\end{figure}
\clearpage

\begin{figure}[!t]
\plotone{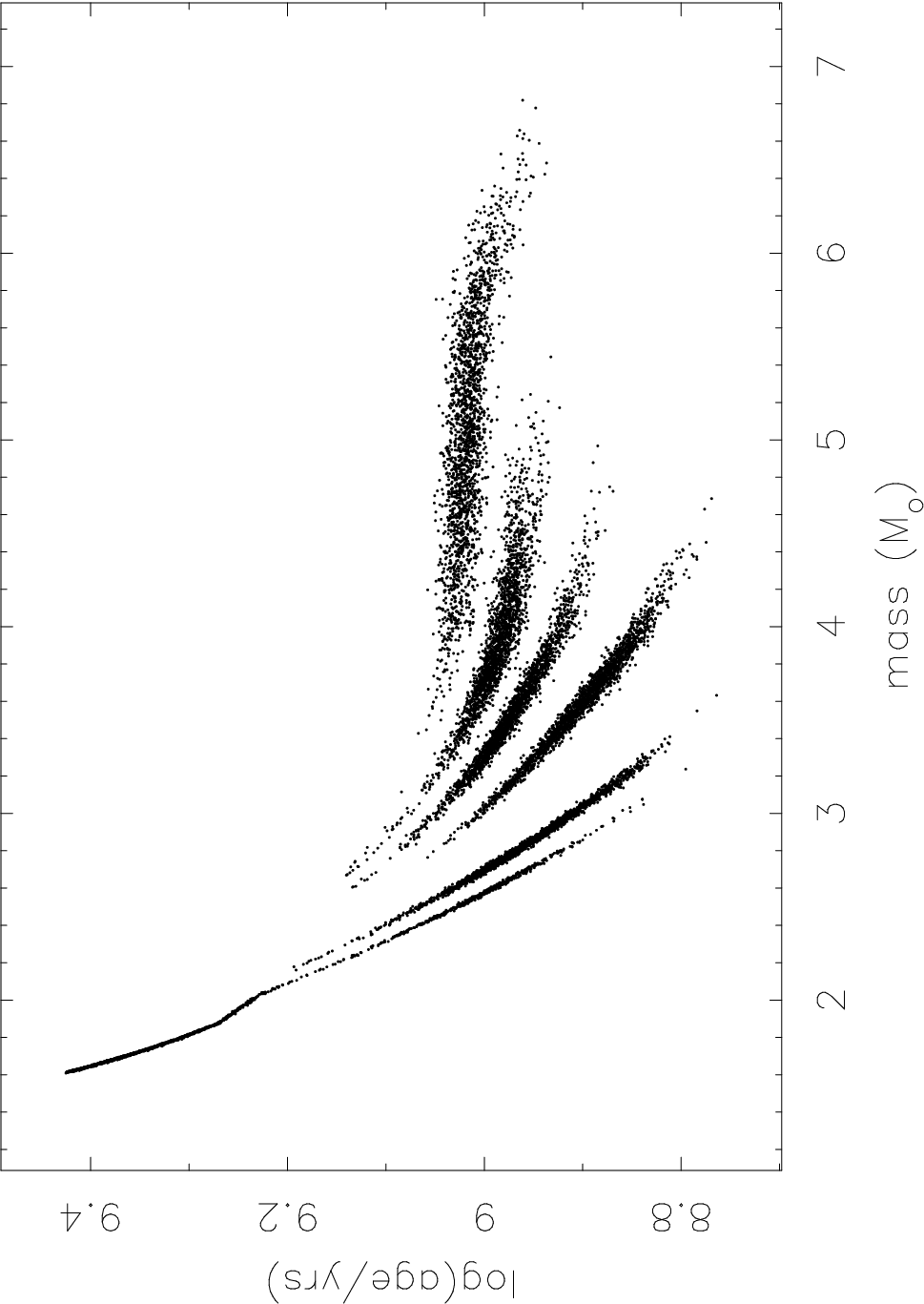}
\centerline{f16.eps}
\end{figure}
\clearpage

\begin{figure}[!t]
\plotone{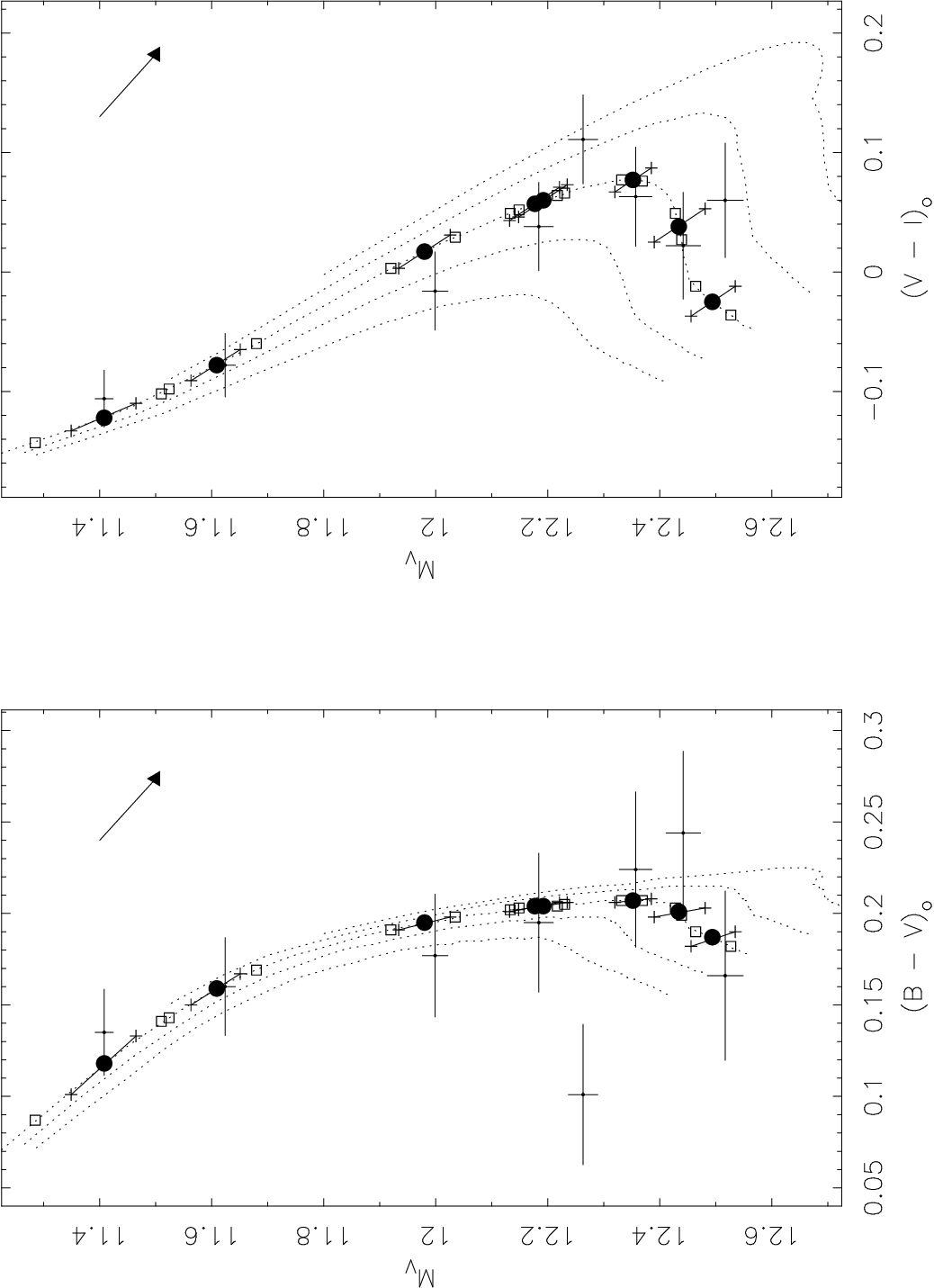}
\centerline{f17.eps}
\end{figure}
\clearpage

\end{document}